\documentstyle[12pt]{article}
\begin{document}

\vspace{1.5ex}
\vspace{4ex}
{ \parindent=120mm {AS-ITP-93-1} }
\parskip 8pt
\begin{center}
{\large \bf THE CONSTRAINT FOR THE LOWEST LANDAU LEVEL AND THE CHERN-SIMONS
FIELD THEORY APPROACH FOR THE FRACTIONAL QUANTUM HALL EFFECT: INFINITE AND
FINITE SYSTEMS \\}

\end{center}
\vspace{6.4ex}
\centerline{\bf Zhong-Shui Ma$^{a,b}$, Zhao-Bin Su$^a$ }
\vspace{4.5ex}
\centerline{\sf  $^a$ Institute of Theoretical Physics, Academia Sinica}
\centerline{\sf  Beijing 100080, China}\vspace{1ex}
\centerline{\sf  $^b$ Zhejiang Institute of Modern Physics,Zhejiang
University }
\centerline{\sf   Hangzhou 310027, China\footnote{\sf Mailing address.}}
\baselineskip=24pt
\vspace{6ex}
\begin{center}
\begin{minipage}{5in}
\centerline{\bf  ABSTRACT}
\vspace{2ex}

{\it We build the constraint that all electrons are in the lowest
Landau level into the Chern-Simons field theory approach for the
fractional quantum Hall system. We show that the constraint can be
transmitted from one hierarchical state to the next. As a result, we
derive in generic the equations of the fractionally charged vortices
( quasi-particles ) for arbitrary hierarchy filling. For a finite
system, we show that the action for each hierarchical state can be
divided into two parts: the surface part provides the action for the
edge excitations while the remaining bulk part is exactly the action
for the next hierarchical states.  In particular, we not only show that
the surface action for the edge excitations would be decoupled from
the bulk at each hierarchy filling, but also derive the explicit
expressions analytically
for the drift velocities of the hierarchical edge
excitations.}

\end{minipage}
\vspace{1.5ex}
\end{center}

{\raggedright {\sf PACS numbers: 73.20.Dx; 73.50.Jt\\}}
\vfill
\eject

{\raggedright{\large \bf I Introduction\\}}

The discovery of the fractional quantum Hall effect (FQHE) [1] has
stimulated extensive studies on the two dimensional quantum
many-electron system in a strong magnetic field.
A considerable progress [2] has been made in understanding for the FQHE
following upon the seminal paper of Laughlin's [3]. The description of
incompressible fluid states of two dimensional electron system in a magnetic
field has provided a key element for such understandings[2,3]. The
analogue of electrons and holes with the fractional charge in a new type
of many body condensates leads to a natural interpretation for the
hierarchy scheme of the FQHE[4]. On the other hand,
motivated by the analogies between the FQHE and the superfluidity [5] as
well as the
existence of large ring exchanges on a large length scale [6], Girvin and
MacDonald [7] raised a subtle question whether there is an off-diagonal
long range order (ODLRO) in the FQHE ground state. They also notice that such
a ODLRO might not have the same physics in the usual sense. By introducing
a 2+1 dimensional bosonization transformation, they did find a sort of the
ODLRO for the bosonized Laughlin wave functions [7,8]. Such an
observation gives rise an
interesting quasi-particle picture that of a charged electron in the presence
of a point ``vortex-tube'' [9]. Since then on a vast number of works appeared
for the field theoretical realization of the fractional quantum
statistics and the
effective field theory description for the FQH system. Among others, the
Ginzburg-Landau Chern-Simons approach (GLCS) [10,11,12] successfully
interpretes a variety of the properties for the FQH system from an
{\it ab intio} point of view.
The chiral Luttinger liquid approach [13,14,15] for the edge excitations [16]
exhibits a deep insight for such an interesting system. And the topological
order approach for the long wave length behavior of the quantum Hall fluid
[17] interpretes a novel sort of the order which is not associated with broken
symmetries but
topological in nature, and it can be characterized by a series of quantum
numbers. Furthermore, the C-S field theory approach for the FQHE can be also
formulated in the fermionic picture which also interpretes various properties
for the FQH system [18].

Despite the successes for the various effective field theory approaches,
we still
have the following questions: (i) whether one should build
in the constraints that all the
electrons are in the lowest Landau level (LLL) from the very beginning of
these approaches. As we have seen in [10,11], the ``trivial Gaussian
fluctuation'' in the GLCS approach arises actually from the inter-Landau
level degrees of freedom. From a more basic point of view, it is
known that the FQH system is essentially a 1+1 dimensional system. The
one dimensional nature of the FQH system should be a direct
consequence of the LLL constraint. (ii) Moreover, different from those
`` conventional ''
vortices, which have their effective mass depending on the mass of the
constituting
particles, we expect that the explicitly built LLL constraint may
play a crucial role for introducing a proper description for the massless
vortices in the hierarchical FQH system in the context of C-S
field-theoretical approach. (iii) A complete C-S field-theoretical approach
for the FQH should not apply only to an infinite FQH system but also to a
finite system. Since the propagation of the `` rippling wave '' along the
boundary for a finite FQH system
is essentially induced by the vortices on the boundary, therefore, if we could
have a correct as well as  unified description for the vortices in the
FQH system, it is natural
to raise the question whether we could have a description for a
finite FQH system in which the action for the edge excitations could be
derived branch by branch from the bulk actions for the corresponding
hierarchical states successively.
And whether the constraint for the LLL would play a non-trivial role again in
such a `` unified '' description.

Motivated by the above arguments, in this paper, we succeed in building
explicitly
the LLL constraint into the C-S field-theoretical description for the FQH
system and show that both the action and the constraint can be transmitted from
one hierarchical state to the next. As its primary
consequence, besides the quantization conditions for the FQHE states as
well as the corresponding hierarchy scheme [4] can be deduced as usual,
the equations for the fractionally charged vortices for any of these
hierarchical levels
can be derived in generic without any mass scale dependent coefficient.
It also does not depend
on whether the FQHE has a BCS type of the symmetry breaking [12]. We can
calculate
accordingly the quasi-particle energy without difficulty. For a finite FQH
system, by applying
a careful treatment of the partial integrations to the actions, we
show that the action for each hierarchical state can be split
into two parts: a surface part provides the action of the edge excitations
and the remaining bulk part is exactly the action for the next hierarchical
states.
In particular, the surface action for the edge excitations
could be decoupled from the bulk only at each hierarchy filling.
Moreover, for the n-th FQH hierarchical states, we
derive analytically the expressions for the drift
velocities for all the $n$ branches of edge excitations which are different
with each other and might be checked in certain properly
designed experiments. To our knowledge, this might
be a first time derivation for the hierarchical expressions for
such drift velocities of the edge excitations.
We thus provide a full dynamical description for
both infinite and finite hierarchical FQH systems. This approach provides
also a field theoretical background for the description of the vortices
in the FQH system
( quasi-particles ) which can have only zero effective mass [19].

Our treatment, in certain sense, is based upon
the Dirac quantization procedure [20] proceeded in the first quantization
representation. It provides a sound background for the treatment for systems
with constraints {\it i.e.}, what we have here is the constraint for the
LLL. If we restrict ourselves only for the first hierarchical level:
the C-S field theory for the bosonized electrons,
we may have almost the same results as those we derived in the following
without the application
of the Dirac quantization method. But it turns out that such a quantization
procedure provides a unified highlight as well as a practically applicable
method for the massless vortices of all the hierarchical states, which are,
in fact,
produced as the singular world lines of the phase variables of the wave
fields hierarchically.

We would try to present our discussions as transparent as possible with
all those detail derivations being properly included. On the meanwhile,
we would like to
expose all the details of our approaches if there is anything inappropriate
even mistaken.

In section II we would treat the constraint for the LLL along the
Dirac algorithm [20] and build it [21,22]
into the dynamical description for the FQH system. Then we apply the
bosonization to the fermion field which makes the bosonized electrons behave
as the singular vortices controlled by the C-S gauge field. We obtain
a complete path-integral description of the FQH system in the context of
2+1 dimensional
C-S field theory, in which the projection to the LLL being carefully
considered. In section III, by introducing the generatized $\rho$ ( particle
density)-$\theta$ (phase variable conjugate to the particle density)
representation [10, 11, 21] for the $Z$-generating functional, we show that
the constraint for the LLL plays a crucial role in the description for the
quasi-particles and, as a result, we provide a generic description for the
quasi-particles of the FQH system which  applies to all
hierarchical states.

Section IV is devoted specially to the finite FQH system which in fact
constitutes one of the main chapters of this paper while sections II and III
might be understood, in certain sense, as the stepping stones for this and the
following sections. In this section, after introducing certain proper
description for the boundary of a finite two dimensional FQH system, we
present a unified
treatment for the surface as well as the bulk degrees of freedom and derive
the action for the edge excitations from the bulk with both actions being
fixed dynamically. It is interesting
to realize that the constraint equation once again plays an essential role
even in the derivation for the surface actions.

Section V actually completes our approach by showing that it really works
for one hierarchical level to the next. We  derive successively the bulk
actions, the equations for the vortices
and edge excitations for the next hierarchical level in detail.
Right on the filling of the second hierarchical level, we show there are
two coexisting branches of edge excitations which couple to each other but
decouple from the bulk system. We distinguish further two limiting cases:
the ``strong coupling'' limit at which the two branches of edge waves
couple to each other strongly and the ``weak coupling'' limit at which
these two branches are further decoupled. Base upon these discussions, we might
conclude that this formalism really provides a hierarchical description
for the finite FQH system. In particular,
we derive the explicit expressions for the propagation velocities of
the edge excitations hierarchically, which should satisfy a sum rule with
interesting physical consequence.

The Appendix A concerns the crucial gauge invariant properties for a finite
FQH system in the context of C-S gauge field approach,
while the Appendix B deals with the decoupling of branches of
edge waves in the weak coupling limit.

All our calculations are given in the nonrelativistic framework.

\bigskip
\bigskip

{\raggedright{\large \bf II The FQH System As A Dynamical
System With The Second Class Constraint \\}}

We consider a two dimensional N-electron system subjected to a strong
perpendicular magnetic field $B$ while all the electrons being in the lowest
Landau level. The Lagrangian for the system has the expression as [6]
$${\cal L}=-{e \over c}\sum_i{\dot {\bf r}_i}\cdot {\bf A}({\bf r}_i(t))
-\sum_{i<j}V({\bf r}_i-{\bf r}_j)
\eqno{(2-1)}$$
where ${\bf r}_i(t)$ is the two dimensional coordinate for the i-th electron
with $i=1,\cdots,N$, ${\dot {\bf r}_i}(t)=d{\bf r}_i(t)/dt$, ${\bf A}(
{\bf r}_i(t))$ is the vector potential for the uniform applied magnetic
field $\bigtriangledown \times {\bf A}=B$ and $V({\bf r}_i-{\bf r}_j)$ is
the interaction between electrons. Throughout this paper, we shall take the
axial gauge as ${\bf A}=$ ( $ -By/2$, $Bx/2$, $0$ )
and the convention that electron's charge equals to $-e$ for
convenience. Different from those ordinary system, the kinetic energy term,
which usually has a bilinear form of the ${\dot {\bf r}}(t)$'s, is absent
in eq. (2-1). Consequently, the canonical momentum ${\bf p}_i$ conjugating to
${\bf r}_i$: $\partial {\cal L}/ \partial
{\dot {\bf r}_i}=-(e/c){\bf A}$, would be independent of ${\dot {\bf r}_i}
(t)$'s. Following the Dirac's algorithm [20], it can be shown that we now have
the second class constraint as
$${\bf \Pi}_i\equiv {\bf p}_i+{e \over c}{\bf A}_i \approx 0
\eqno{(2-2)}$$
where $\approx $ indicates Dirac's weak equality [20], and then the N-electron
Hamiltonian for the system takes the form as
$${\cal H}=\sum_{i<j} V({\bf r}_i-{\bf r}_j)
\eqno{(2-3)}$$
Moreover, the canonical quantization for a system with constraints could be
accomplished by the correspondence principle as: to replace the Dirac bracket
$\{,\}_D$ of any couple of dynamical variables $f$ and $g$, {\it i.e.},
$\{f,g\}_D$, by a quantum commutator $[f,g]/i\hbar$, where $[f,g]\equiv
fg-gf$ and the canonically invariant Dirac bracket is defined as
$$\{ f, g \}_D=\{f,g\}-\sum_{\alpha,\beta;i,j}\{f,{\Pi^\alpha}_i\}{C_{\alpha,
\beta}}^{-1}\{{\Pi^\beta}_j,g\}
\eqno{(2-4)}$$
In eq. (2-4) the script brackets without the subscript $D$ are the usual
Poisson brackets and $\alpha$, $\beta$ are the scripts for the 2-dimensional
vector components. The matrix elements of $C$ are given by $C_{\alpha,
\beta}~[i,j]\equiv
\{\Pi_i^\alpha,\Pi_j^\beta\}$ and ${C_{\alpha \beta}}^{-1}\equiv
(C^{-1})_{\alpha \beta}$. We notice further that
$$\{{\Pi^\alpha}_i,{\Pi^\beta}_j\}=-\in_{\alpha \beta}\delta_{ij}{\hbar \over
\lambda^2}
\eqno{(2-5)}$$
where the second rank antisymmetric tensor is defined as $\in_{12}=
-\in_{21}=1 $ and the magnetic length $\lambda=(\hbar c/eB)^{1\over 2}$.
As a result, $C_{\alpha \beta}~[i,j]$ is a non-singular matrix. We may then
work out all the Dirac brackets of the canonical variables and further quantize
them. The only nontrivial commutation relation is found as
$$[x^\alpha_i,x^\beta_j ]=i\in_{\alpha \beta}\delta_{ij}\lambda^2
\eqno{(2-6)}$$
{\it i.e.}, the application of the Dirac quantization procedure
to the system that
all electrons are in the LLL makes the electrons' coordinates acquire the
physics of their guiding center coordinates while the canonical momentum being
consistently eliminated {\it via} the Dirac brackets.
We may verify without difficulty that the constraint for the LLL can be
equivalently described by the following
constraint for the N-electron wave function defined in the conventional
2-dimensional space as
$$\Pi_i\Psi ({\bf r}_1,\cdots ,{\bf r}_N)=0
\eqno{(2-7)}$$
together with the understanding that, not only the real processes, but also
all the
virtual processes beyond the subspace of eq. (2-7) are prohibited at all,
where $\Pi_i=(\Pi^x_i-i\Pi^y_i)/{\sqrt 2}$. A detail account for the
application
of the Dirac's quantization on such a constraint system is presented in
literature [22].

Base upon the above treatment which is accomplished in the first quantization
representation, we may introduce the corresponding description in the second
quantization representation accordingly. Following eqs. (2-3) and (2-7), the
second quantized Hamiltonian now has the form as
$$H=V[{\hat\Psi}^+(x){\hat\Psi}(x)-\rho_{BG}]$$
$$={1 \over 2}\int d^2r_1
d^2r_2 ({\hat\Psi}^+({\bf
r}_i){\hat\Psi} ({\bf  r}_1) -\rho_{BG})V({\bf r}_1-{\bf r}_2)({\hat\Psi}^+
({\bf r}_2)
{\hat\Psi} ({\bf r}_2)-\rho_{BG})
\eqno{(2-8)}$$
while the electron wave field operator ${\hat\Psi} ({\bf r})$ satisfying the
fermion statistics is subjected to a LLL constraint that
$$\Pi {\hat\Psi} ({\bf r})=0
\eqno{(2-9)}$$
where $\rho_{BG}\equiv S^{-1}\int d^2r \Psi^+({\bf r})\Psi ({\bf r})$
with $S$ being the total area of the system and
should be equal to the average charge density contributed by
the positive background. One can easily verify that the projection to the
LLL, even for the virtual processes, is rigorously guaranteed by the
constraint (2-9) in the second quantization representation.

By applying the standard procedure, now we introduce further the  bosonized
representation $\Phi (x)$ for the electron field $\Psi (x)$ [7,10,11] as
$$\Psi (x) =e^{i\Theta (x)}\Phi (x)
\eqno{(2-10)}$$
with the definition
$$\Theta (x)=m \int d^2z' Im \ln ({\bar z}-{\bar z'}) \rho (z')
\eqno{(2-11)}$$
and the C-S gauge field can be defined as
$${\bf a}(x)=\bigtriangledown \Theta (x)
\eqno{(2-12)}$$
In eq. (2-11) and the following, it is often convenient to introduce the
complex notations as
$$z={1 \over {\sqrt 2}}(x+iy), {\bar z}={1 \over {\sqrt 2}}(x-iy)$$
$$\partial ={\partial \over {\partial z}}={1 \over {\sqrt 2}}({\partial
\over \partial x}-i{\partial \over \partial y}),{\bar \partial }={\partial
\over \partial {\bar z}}={1 \over {\sqrt 2}}({\partial \over \partial x}
+i{\partial \over \partial y})
\eqno{(2-13)}$$
and
$${\bf A}={1 \over {\sqrt 2}}(A_x-iA_y)=-i{B \over 2}{\bar z},
{\bar {\bf A}}={1 \over {\sqrt 2}}(A_x+iA_y)=i{B \over 2}z
\eqno{(2-14)}$$
Substituting eqs. (2-10), (2-11) and (2-12) into eqs. (2-8) and (2-9), and
noticing eqs. (2-13) and (2-14), we have
$$H={1 \over 2}\int d^2r_1 d^2r_2 ({\hat\rho} ({\bf r}_1)-\rho_{BG})V
({\bf r}_1-{\bf r}_2)
({\hat\rho} ({\bf r}_2)-\rho_{BG})
\eqno{(2-15)}$$
and the LLL constraint becomes
$${\tilde\Pi}{\hat\Phi}({\bf r}) \equiv ({\partial \over \partial z}+i{1 \over
{\lambda^2 B}}A+ia){\hat\Phi} ({\bf r})=0
\eqno{(2-16)}$$
In eq. (2-15), ${\hat\rho} (z)={\hat\Phi}^+ (x){\hat\Phi} (x)$ and
$\rho_{BG}=S^{-1}\int
d^2r \Phi^+ ({\bf r})\Phi ({\bf r})$. Due to the singular behavior of function
$Im\ln ({\bar z}-{\bar z}')$, following from eq. (2-11),
we may derive
$$\in_{\alpha \beta}\partial_\alpha a_\beta =i({\bar \partial }a-\partial
{\bar a})=-2\pi m \rho ({\bf r})
\eqno{(2-17)}$$
which relates the `` magnetic field " of
`` C-S gauge potential " $a_\alpha$ to the particle density and has
the physical intuition as: attaching m-`` magnetic '' flux of the C-S field
to an electron [10,11].
If we  impose further the equation of continuity, ${\dot \rho}({\bf r})+
\partial_\alpha j_\alpha ({\bf r})=0$, then, the time derivative
of `` C-S gauge
potential " should relate to the matter current as
$$\in_{\alpha \beta}{\dot a}_\beta=2\pi m j_\alpha ({\bf r})+\in_{\alpha
\beta} \partial_\beta a_0
\eqno{(2-18)}$$
up to a trivial divergence free term.

Taking into account of all the above considerations as well as the fact that
the constraint for the LLL should be imposed on all the time slices in the
dynamical
evolution, the path integral representation for the Z-generating functional
would have the following form
$$Z[A]=\int {\cal D}\Phi {\cal D}\Phi^+ {\cal D}a_\mu \delta [{\tilde \Pi}
\Phi ]\delta [\Phi^+ {\tilde \Pi}^+]\exp\; (i\int d^3x {\cal L}_0)
\eqno{(2-19)}$$
with
$${\cal L}_0=\Phi^+(i\partial_0-a_0)\Phi -V[\rho -\rho_{BG}]
-{1 \over 2\pi m}a_0\in_{\alpha \beta}\partial_\alpha a_\beta+{1 \over
4\pi m}\in_{\alpha \beta}a_\alpha {\dot a}_\beta
\eqno{(2-20)}$$
where the gauge fixing condition is understood involved implicitly and $\delta
[\cdots]$ is the $\delta$-functional. Comparing to the conventional 2+1
dimensional C-S field theory, we have not only two second class
constraints for the LLL being
explicitly built in but also an action in which the kinetic energy is
absent. In fact this is a sort of the non-relativistic C-S field theory
with its interacting matter field being massless.

\bigskip
\bigskip
{\raggedright {\large \bf III Description For The Vortices ( Quasi-particles )
In The FQH System\\}}

Since now we
are in the boson representation, we prefer to introduce the phase $\theta (x)$
and the electron density $\rho (x)$ for the wave field as the dynamical
variables by taking
$$\Phi (x)={\sqrt {\rho (x)}}e^{i\theta (x)}
\eqno{(3-1)}$$
The phase variable $\theta (x)$ bears the description for the vortices and can
be
further decomposed into a regular part $\theta_r$ and a singular part
$\theta_s$ as [10,11]
$$\theta (x)=\theta_r (x)+\theta_s (x)
\eqno{(3-2)}$$
in which $\theta_r$ and $\theta_s$ satisfy
$$\in_{\alpha \beta}\partial_\alpha \partial_\beta \theta_r=0
\eqno{(3-3)}$$
and
$$\in_{\alpha \beta}\partial_\alpha
\partial_\beta \theta_s=-2\pi\rho_s(x)
\eqno{(3-4)}$$
respectively. We notice that $\rho_s$ has
the physical intuition as the density for the vortices.
We then substitute eq.(3-1) into eq.(2-16) and its conjugate,
the constraint for the LLL can then be expressed in terms of $\rho$-$\theta$
variables as
$$ f[\rho, \theta]\equiv ({1\over 2}{{\partial \ln \rho}\over
{\partial z}}+i{{\partial \theta_r}\over
{\partial z}}+i{{\partial \theta_s}\over {\partial z}}+i{1 \over \lambda^2B}
A+ia)=0$$
$$ f^*[\rho, \theta]\equiv ({1\over 2}{{\partial \ln \rho}\over {\partial
{\bar z}}}-i{{\partial \theta_r}\over {\partial {\bar z}}}-i{{\partial
\theta_s}
\over {\partial {\bar z}}}
-i{1 \over \lambda^2B}{\bar A}-i{\bar a})=0
\eqno{(3-5)}$$

The $Z$-generating functional (2-19) becomes
$$Z[A]=\int {\cal D}\rho {\cal D}\theta_r {\cal D}\theta_s {\cal D}a_\mu
\delta [f[\rho,\theta]]\delta [f^*[\rho,\theta]]$$
$$\exp~~i\int d^3x\{\rho (-{\dot \theta}_r-{\dot \theta}_s-a_0+e\varphi)-
V[\rho-{\bar \rho}]-{1 \over 2\pi m}\in_{\alpha \beta}a_0 \partial_\alpha
a_\beta+{1 \over 4\pi m}\in_{\alpha \beta}a_\alpha {\dot a}_\beta\}
\eqno{(3-6)}$$
where we included an applied electric field with ${\varphi (x)}$ being
its scalar potential.
It is quite
clear from eqs. (3-5) and (3-6) that, as a result of introducing the
$\rho$-$\theta$ representation, the C-S field acquires a gauge term: $a_\mu
\to a_\mu +\partial_\mu \theta_r$, $\mu=0,1,2$, not only in the matter part
of the action but also in the constraints. It is known that the action for
the C-S term of the gauge field itself is
invariant respect to the local gauge transformation up to a surface term.
Therefore, we may eliminate the regular part of the phase variables
$\theta_r$ by performing a gauge transformation $a_\mu \to a_\mu-\partial_\mu
\theta_r$ for the Z-generating functional expression eq. (3-6) and forget
about the induced surface term ${\cal K}_\Gamma[a,\theta_r]$ tentatively. We
will come back to this induced surface term in the next section.
Moreover, by taking a linear combination of $\partial f^*/\partial z$ and
$\partial f/\partial {\bar z}$ in which the $\theta_r$ has been eliminated
as just mentioned, the constraints eq. (3-5) can be transformed into the
following equivalent form as
$${1 \over 2}\bigtriangledown^2\ln \rho +{1 \over \lambda^2}-2\pi \rho_s
+\in_{\alpha \beta}\partial_\alpha a_\beta =0
\eqno{(3-7)}$$
and
$$\bigtriangledown \cdot {\bf a}=0
\eqno{(3-8)}$$
We then carry out the integration over the zero-component C-S field $a_0$
in eq. (3-6) and
recover the C-S constraint (2-17) first. By solving eqs. (3-8) and (2-17),
we may integrate further ${\cal D}a_1{\cal D}a_2$ in eq. (3-6). Finally we
derive
$$Z[A]=\int {\cal D}\rho {\cal D}\theta_s \delta[{\cal F}[\rho, \theta_s;
B]]\exp~~i\int d^3x \{-\rho{\dot \theta}_s+e\rho\varphi -V[\rho-{\bar \rho}]
+{1\over 4\pi m}\in_{\alpha \beta}a_\alpha {\dot a}_\beta\}
\eqno{(3-9)}$$
with
$${\cal F}[\rho,\theta_s;B]\equiv {1\over 2}\bigtriangledown^2\ln\rho+{1
\over \lambda^2}-2\pi m\rho -2\pi \rho_s=0
\eqno{(3-10)}$$
and $a_\alpha$ being now the solution of eq.(2-17) in consistency with the
gauge fixing condition eq. (3-8). In this equation, the term $\lambda^{-2}$
could be understood as $(e/\hbar c)\bigtriangledown \times {\bf A}$. We would
like to emphasize
here that apart from surface term ${\cal K}_\Gamma[a,\theta_r]$ contributed
by the C-S term due to the
gauge transformation $a_\mu \to a_\mu -\partial_\mu \theta_r$, we have not
done any partial integration in the above derivations.

By now we derive the Z-generating functional for the FQH system in the
$\rho$-$\theta$ representation.
We see that the LLL constraint not only makes the electrons'
kinetic energy disappear, but also manifests itself as a functional
relation among $\rho$, $\rho_s$ and $B$: ${\cal F}[\rho,\rho_s;B]=0$,
which plays a crucial role in the understanding of the properties for the
FQHE states. The contributions from the C-S field which had been introduced
non-trivially for the bosonization procedure now transfer partly their
effect to the
statistics index ``$m^{-1}$'' appearing in the constraint functional
${\cal F}[\rho, \rho_s;B]$ while the remaining effect is still born by the
term $(4\pi m)^{-1}\in_{\alpha \beta}a_\alpha {\dot a}_\beta$. If we imagine
the functional integral ${\cal D}
\rho$ in eq. (3-9)
being carried out, we may understand that the eq. (3-9) describes
a system with $\rho_s$ as its only independent dynamical variable.
Since $\in_{\alpha \beta} \partial_\alpha \partial_\beta$ can be nonzero
only at certain singular 2+1 dimensional world lines, so $\theta_s$ is a
smooth functional in space except those singular points (at vortex
positions). We interprete these propagating singular points as point
particle-like vortex cores. Then
the vortex  density should have the expression as $\rho_s ({\bf x})=\sum_j q_j
\delta^2 ({\bf x}-{\bf x}_j (t))$ with $q_j=\pm 1$ being the vortex charge and
${\bf x}_j(t)$'s being the world line
for the j-th vortex. The vortex current $j_s^\alpha ({\bf x})=\sum_jq_j
{\dot{\bf x}}_j^\alpha(t) \delta^2 ({\bf x}-{\bf x}_j(t))$ can also be
equivalently expressed as
$$j_s^\alpha ({\bf x})={1 \over 2\pi}\in_{\alpha \beta}(\partial_0
\partial_\beta-\partial_\beta \partial_0 )\theta_s
\eqno{(3-11)}$$
We can easily verify that the expressions (3-4) and (3-11) are consistent
with the conservation of the vortex current: ${\dot \rho}_s+\partial_\alpha
j_s^
{\alpha}=0$.
Kept with the above understandings, it is obvious that in the expression
for the Z-generating functional eq. (3-9), the path integral over
${\cal D}\theta_s$ is essentially an evolution in the first quantization
representation for the vortices.

It is straightforward to derive from the Z-generating functional eq. (3-9)
the following equation
$${1 \over 2}\bigtriangledown^2 <\ln\rho >-2\pi m <\rho >+{1 \over
\lambda^2}-2\pi <\rho_s>=0
\eqno{(3-12)}$$
where $<\cdots>$ is the path integral average over the normalized Z-generating
 functional, {\it i.e.}, average over the physical ground state. This equation
in fact had been first time derived directly from the constraint equations
for the LLL by applying the collective field theory
approach [21,23]. What we have here more is to make its connection to the
dynamics being explicit. For a homogeneous system with zero vortex, we
derive the quantization
condition from eq. (3-12) for the FQHE states, ${\bar \rho}
=(2\pi m\lambda^2)^{-1}$, immediately. For a single vortex, we can draw the
conclusion easily
from this equation that it carries a fractional charge of $qe/m$
where $q>0$ corresponds a quasi-hole. So this
equation can be interpreted as the equation for the vortices ( quasi-particle )
of the first hierarchy. Its mean field solution can be solved numerically
without difficulty and then the energy for the quasi-particles
can be calculated subsequently. We notice that different from the usual G-L
type description,
there is no mass-scale dependent parameter appearing
in eq. (3-12). It also does
not depend on whether there is a `` BCS type symmetry breaking '' [12] in the
FQHE state.

In the constraint equation (3-10), $\rho_s$ has the $\delta$-function like
singularities at the location of each vortex. While the main role played by
the $\bigtriangledown^2\ln \rho$ is to cancel such singularities since the
$\rho
({\bf r})$ should have certain drastic variations close to the vortex centers.
If we further introduce the second quantization representation for the
vortices, such singularities would be smeared out in the wave field
description.
Hence the $\bigtriangledown^2 \ln \rho$ term would be no more interesting as
the main physics are usually controlled by the long wave
length behaviors.
Therefore, for sake of convenience, we would ignore the $\bigtriangledown^2
\ln \rho$ term in the following with the understanding that  there is always a
term $-\in_{\alpha \beta}\partial_\beta \ln \rho /2$ associated with
$\partial_\alpha \theta_s$ implicitly in the
first quantization representation of the vortices, while such a term could
be reasonably ignored in its second quantization representation.
\bigskip
\bigskip

{\raggedright {\large \bf IV Intimate Relation Between Edge Excitations
And Hierarchical Structure For A Finite FQH System \\}}

Now we shall treat the finite FQH system, {\it i.e.}, to separate the
surface part of the action properly from the bulk part for
a finite FQH system. Before going into the details we would like to introduce
certain descriptions for the boundary of a finite FQH system. We imagine
that the two dimensional system is enclosed by a (spatially) one dimensional
boundary $\Gamma$. The continuity equation ${\dot \rho}+\partial_\alpha
j_\alpha =0$ can then be written in the integral form as
$$\int d^2 x \partial_t \rho=-\oint _\Gamma dl~~ n_\alpha \rho v_\alpha
\eqno{(4-1)}$$
where $dl$ is the linear integral along the boundary and $n_\alpha$ is the unit
normal vector of the boundary being defined always oriented outward from the
system. If we imagine
a finite period of time $\delta t$, it becomes $\int d^2x\delta \rho=-
\oint_\Gamma dl~ n_\alpha \rho \delta r_\alpha$ in which
we have introduced a displacement vector $\delta {\bf r}$ defined formally
along the boundary. We may express $\delta \rho$ as
$\delta \rho =\rho -{\tilde \rho}$,
where $\tilde \rho$ is certain initial distribution of the electrons in the
system. Then, we have
$$\int d^2x (\rho -{\tilde \rho})=-{\tilde \rho}\oint_\Gamma dl~~n_\alpha
\delta r_\alpha
\eqno{(4-2)}$$
If we take ${\tilde \rho}={\bar \rho}$ with ${\bar \rho}$ being the average
electron density, the
lefthand side of the equation should be zero, so that we should have
$$\oint_\Gamma dl~~n_\alpha \delta r_\alpha=0
\eqno{(4-3)}$$
Consequently $\delta r_\alpha$
can be interpreted either as the displacement for the particles (electrons)
passing back and forth through the boundary or as the ``rippling''
displacement for the boundary [15] deviating out- or inward along the boundary.
Obviously, it is understood that these equations are valid up to the first
order of $\delta {\bf r}$. If we split $\theta_s$ into two parts:
$\theta_s=\theta_s^{bulk}+\theta_s^{surf}$, correspondingly,
$$\rho_s=\rho_s^{bulk}+\rho_s^{surf}
\eqno{(4-4)}$$
we then have
$$\rho_s^{bulk}=- {\frac 1 {2\pi}}
\in_{\alpha \beta}
\partial_\alpha \partial_\beta\theta_s^{bulk}
\eqno{(4-5)}$$
which contributes to the average vortex density of the system ${\bar \rho}_s$
and
$$\rho_s^{surf}=- {\frac 1 {2\pi}}
\in_{\alpha \beta}\partial_\alpha \partial_\beta
\theta_s^{surf}
\eqno{(4-6)}$$
which is nonzero only at the boundary, and has zero contribution to the
${\bar \rho}_s$ so that ${\bar \rho}_s ={\bar \rho}_s^{bulk}$. Making use
of the constraint eq. (3-10), we can have both
$$\rho={1\over 2\pi m \lambda^2}-{1 \over m}(\rho_s^{surf}+\rho_s^{bulk})
\eqno{(4-7)}$$
and
$${\bar \rho}={1 \over 2\pi m\lambda^2}-{1\over m}{\bar \rho}_s^{bulk}
\eqno{(4-8)}$$
where the $\bigtriangledown^2 \ln \rho$ terms are ignored with the previously
mentioned understanding. By taking ${\tilde \rho}={\bar \rho}$ and then
substituting eqs. (4-7) and (4-8) into eq. (4-2), we may draw the expression
for $\delta {\bf r}$ from eq. (4-2) as
$$\delta r_\alpha =-{1 \over 2\pi m{\bar \rho}}\in_{\alpha \beta}\partial_\beta
\theta_s^{surf}
\eqno{(4-9)}$$
up to an arbitrary gauge transformation $\theta_s^{surf}\to \theta_s^{surf}
+\theta'_s$ where $\theta'_s$ is a regular function defined along the
$\Gamma$: $\oint_\Gamma dl n_\alpha \in_{\alpha \beta}\partial_\beta
\theta'_r=0$ but not determined yet.

Moreover, since a finite two dimensional FQH system is always confined by
some potential, its chemical potential, $\mu$, is determined
in such a way that the Gibbs free energy is minimized consistently with
the spatial distribution of the electrons. Therefore, the local deviation of
the applied electric potential, $e\varphi$, from the chemical potential at
the boundary is equal to the work done by those electrons that passed
through the
boundary, or in another words, due to the local displacement of the boundary
from its equilibrium configuration. Again in the sense of the first order
deviation, we should then have
$$(e\varphi -\mu )|_\Gamma =e(\varphi-\varphi_0)|_\Gamma=-e{\bf E}\cdot \delta
{\bf r} |_\Gamma
\eqno{(4-10)}$$
where ${\bf E}$ is the applied electric field and can be expressed as
${\bf E}=-{\bf \bigtriangledown} \varphi $.

Intuitively, the boundary is an ``infinitesimally"
thin layer with a "thickness" of order of the "rippling" displacement
$\delta {\bf r}$. Such a boundary layer
is a layer of $\rho_s^{surf}$, $i.e.$, in which and only in which
$\rho_s^{surf}$ has nonzero value locally. It has further the following
properties
$$\int_{x\subset \Gamma} d^2 x {\rho_s}^{surf}=0
\eqno{(4-11)}$$
and
$${\bf j}_s^{bulk} \cdot {\bf n} |_{x\subset \Gamma}=0
\eqno{(4-12)}$$
where $\int_{x\subset \Gamma}d^2x$ means a 2D integration carring over only
this surface layer region.
We may also verify without difficulty that
eqs. (4-11) and (4-12) are consistent with eqs. (4-4) to (4-8).
Eq.(4-11) has the physical meaning similar to those of
$\delta {\bf r} \cdot {\bf n}$ in eq.(4-3) that
$\rho_s^{surf}$ describes the local accumulation or dissipation of the
particles in the surface layer with
its total accumulation (dissipation)
being kept equal to zero.
Moreover, since the description for the displacement of the
particles (electrons) passing back and forth through the boundary (which
results the local accumulation and dissipation of the particle density)
has been taken care by eq.(4-11), as a result, we should have eq.(4-12)
for consistency. We notice also that eq.(4-12) is valid only up to the
leading order where the unit vector ${\bf n}$ is defined as the normal of
the outer boundary of the layer.
If we view the boundary as a surface layer in sense of eqs.(4-11) and (4-12),
then
we can show that
eq. (4-9) applies locally to the whole boundary layer region. In fact,
we may divide imaginary the surface layer further into many sub-layers with
the requirement that each of them having eq.(4-11) being satisfied.
But for now, instead over the whole boundary region, we should have the 2D
integration in eq. (4-11) carrying over only those sub-layers
under consideration. Therefore each
intersurface between two successive sub-layers encloses an area with its
interior bulk part
coinciding exactly with that of the original system but its surface layer
being only an inner part of that of the original system. Obviously we then
can apply the same arguments to derive eq. (4-9) like equation on each
intersurface  in the interior of the boundary layer, so that, eq.(4-9) is
indeed valid within the boundary layer locally. Furthermore,
following the similar
spirit, it is not difficult to verify that eq. (4-10)
is also valid within the boundary layer.

For the term $\int d^2x dt \rho (e\varphi -\mu)$ in the action of eq. (3-9),
by utilizing the constraint eq. (3-10) or eq. (4-7), we have
$$\int d^2x dt \rho (e\varphi -\mu)=\int d^2x dt [{1 \over {2\pi m\lambda^2}}
-{1 \over m}(\rho_s^{surf}+\rho_s^{bulk}) ](e \varphi -\mu)
\eqno{(4-13)}$$
We notice that the term $\int
d^2x dt (2\pi m\lambda^2)^{-1}(e\varphi -\mu )$ in the r.h.s. of the above
equation will not contribute to the dynamics of the system since $e\varphi$
is due to the applied electric potential and $\mu$ is a constant determined
by the envelope potential. We would like further to keep the $\rho_s^{bulk}$
term in the r.h.s. of eq. (4-13) to be retained.
Moreover, by applying eq. (4-6) to the
$\rho_s^{surf}$ which is nonzero only in the boundary
layer, the remaining term in the r.h.s. of eq. (4-13) can be rewritten as
$${1 \over 2\pi m}\int_{x\subset \Gamma} d^2x dt [
\in_{\alpha \beta}\partial_\alpha \partial_\beta
\theta_s^{surf}](e\varphi -\mu)
\eqno{(4-14)}$$
Taking into account of eqs. (4-9) and (4-10) with the understanding that
both of the two being valid in the whole boundary layer, eq. (4-14) becomes
$${1\over {(2\pi m)^2 {\bar \rho}}}\int_{x\subset \Gamma} d^2x dt
(\in_{\alpha \beta}\partial_\alpha \partial_\beta \theta_s^{surf})
(E_\alpha \in_{\alpha \beta}\partial_\beta \theta_s^{surf})
\eqno{(4-15)}$$
We now introduce the following identity for the integrand of the expression
eq. (4-15) as
$$\partial_\alpha M_\alpha\cdot E_\beta M_\beta
\equiv
\partial_\alpha (M_\alpha
E_\beta M_\beta)-{1 \over 2}E_\alpha \partial_\alpha (M_\beta M_\beta)-(\in_
{\alpha \beta}M_\alpha E_\beta)\cdot (\in_{\alpha' \beta'}\partial_{\alpha'}
M_{\beta'})$$
with $M_\alpha$ being identified as $\in_{\alpha \beta}\partial_\beta \theta_s
^{surf}$. Since $\in_{\alpha \beta}\partial_\alpha M_\beta=-\partial_\alpha
\partial_\alpha\theta_s^{surf}$, we may choose the gauge for $\theta_s^{surf}$
and make the last term on the r.h.s. of the above identity becomes zero.
Substituting the identity into expression (4-15) and then
$\int_{x \subset\Gamma}d^2x$ can be transformed into a `` surface ''
integral $\oint_B dl$ which
encloses the boundary layer by two line integral one for the outer boundary
and the other for the inner boundary, {\it i.e.},
$${1 \over {(2\pi m)^2{\bar \rho}}}\int dt\oint _B dl[n_\alpha in_{\alpha
\beta}
\partial_\beta \theta_s^{surf})(E_{\alpha '}\in_{\alpha' \beta'}\partial_
{\beta'}\theta_s^{surf})-{1\over 2}(E_\alpha n_\alpha)(\partial_\beta \theta_s
^{surf}\partial_\beta \theta_s^{surf})]
\eqno{(4-16)}$$
Without lost of generality, we may assume reasonably that up to the leading
order of $\delta r$, $n_\alpha \in_{\alpha \beta}\partial_\beta
\theta_s^{surf}$
being zero at the inner boundary line while $n_\alpha \partial_\alpha \theta_s
^{surf}$ taking the same value locally at the both boundary lines. Noticing
further that $(\partial_\alpha \theta_s^{surf})^2=(n_\alpha \partial_\alpha
\theta_s^{surf})^2+(n_\alpha \in_{\alpha \beta}\partial_\beta \theta_s^{surf})
^2$, then expression (4-14), {\it i.e.}, eq. (4-16) can be transformed into
the following form
$${1 \over {2(2\pi m)^2{\bar \rho}}}\int dt \oint _\Gamma dl (n_\alpha \in_
{\alpha \beta}\partial_\beta\theta_s^{surf})\cdot (E_\alpha \in_{\alpha \beta}
\partial_\beta \theta_s^{surf})
\eqno{(4-17)}$$
Taking into all the above considerations, we derive from eq. (4-13) that
$$\int d^2xdt \rho (e\varphi -\mu )$$
$$={eE\over {2(2\pi m)^2{\bar \rho}}}
\int dt \oint_\Gamma dl (n_\alpha \in_{\alpha \beta}\partial_\beta
\theta_s^{surf})^2-{1 \over m}\int d^2x dt \rho_s^{bulk}(e\varphi -\mu)
\eqno{(4-18)}$$
where we have assumed the electric field always parallel to the normal on the
boundary.

For the first as well as the last term of the action (see eq.(3-9)),
$-\rho {\dot \theta}_s+(4\pi m)^{-1}\in_{\alpha \beta}a_\alpha {\dot a}_\beta$,
we notice $a_\alpha$ is the solution of eq. (2-17) which can be expressed
in terms of $\theta_s$ by making use of eqs. (4-7) and (3-4) as
$$a_\alpha =-\partial_\alpha \theta_s -{1 \over {\lambda^2 B}}A^{em}_\alpha
\eqno{(4-19)}$$
Therefore, by applying further eqs. (4-7) and (4-19)
$$\int d^2xdt (-\rho {\dot \theta}_s+{1 \over {4\pi m}}\in_{\alpha \beta}
a_\alpha {\dot a}_\beta)$$
$$=\int d^2xdt [-{1 \over 2\pi m}(\in_{\alpha \beta}\partial_\alpha \partial_
\beta \theta_s){\dot \theta}_s+{1 \over 4\pi m}\in_{\alpha \beta}\partial_
\alpha \theta_s\partial_0 \partial_\beta \theta_s]
\eqno{(4-20)}$$
where ( and afterward ) we have ignored ( would ignore ) all those integrands
of a total time derivative.
Taking a partial integration with respect to the
`` $\partial_\alpha$ '' in the first term, expression (4-20) becomes
$$-{1 \over 2\pi m}\int dt \oint_\Gamma dl (n_\alpha \in_{\alpha \beta}
\partial_\beta\theta_s){\dot \theta}_s$$
$$+\int d^2x dt [{1 \over 2\pi m}\in_{\alpha \beta}\partial_\beta \theta_s
(\partial_\alpha {\dot \theta}_s-\partial_0\partial_\alpha \theta_s)-{1 \over
4\pi m}\in_{\alpha \beta}\partial_\alpha \theta_s \partial_0\partial_\beta
\theta_s]
\eqno{(4-21)}$$
For the purpose of separating the `` surface '' and `` bulk '' degrees of
freedom, we express $\theta_s$ further as $\theta_s=\theta_s^{surf}+\theta_s
^{bulk}$ in eq. (4-21). Utilizing the following equalities
$$\int d^2x dt \in_{\alpha \beta}\partial_\alpha \theta_s^{surf}\partial_0
\partial_\beta\theta_s^{bulk}=\int d^2x dt \in_{\alpha \beta}\partial_\alpha
\theta_s^{bulk}\partial_0\partial_\beta \theta_s^{surf}$$
$$\partial_0\rho_s^{bulk}+\partial_\alpha j_{s,\alpha}^{bulk}=0$$
$$\partial_0\rho_s^{surf}+\partial_\alpha j_{s,\alpha}^{surf}=0$$
and eq.(4-12), we can derive the following
expression from eq. (4-21) by straightforward calculations
$${1 \over m}\int d^2x dt \partial_\alpha \theta_s^{bulk}j_{s,\alpha}^{bulk}
-{1\over 4\pi m}\int d^2x dt \in_{\alpha \beta}\partial_\alpha
\theta_s^{bulk}\partial_0\partial_\beta \theta_s^{bulk}$$
$$-{1 \over 4\pi m}\int dt \oint_\Gamma dl n_\alpha \in_{\alpha \beta}
\partial_\beta \theta_s {\dot \theta}_s-{1 \over 4\pi m}\int dt \oint _\Gamma
n_\alpha \in_{\alpha \beta}\partial_\beta \theta_s^{bulk}{\dot \theta}_s^{bulk}
\eqno{(4-22)}$$
where we have also utilized the expression for $\rho_s^{bulk,~surf}$ and
${\bf j}_s^{bulk,~surf}$ given by eqs. (4-5), (4-6) and (3-11).

Now we introduce a dual gauge field for the bulk system as
$$A'_\alpha=-{1 \over m}\partial_\alpha \theta_s^{bulk}-{1 \over m\lambda^2B}
A_\alpha^{em}
\eqno{(4-23)}$$
Making use further of eqs.(4-5) and (4-7), it satisfies
$$\in_{\alpha \beta}\partial_\beta A'_\beta=-2\pi \rho^{bulk}
\eqno{(4-24)}$$
Substituting eq. (4-23) into the first two terms of expression (4-22), we
derive step by step the following expression as
$$\int d^2x dt (-\rho {\dot \theta}_s+{1 \over 4\pi m}\in_{\alpha \beta}
a_\alpha {\dot a}_\beta)$$
$$=-{1 \over 4\pi m}\int dt \oint_\Gamma dl
n_\alpha \in_{\alpha \beta}
( \partial_\beta \theta_s {\dot \theta}_s +
\partial_\beta \theta_s^{bulk}{\dot \theta}_s
^{bulk})$$
$$+\int d^2x dt \{-{\bf j}_s^{bulk}\cdot {\bf A}'-{m \over 4\pi}\in_{\alpha
\beta}A'_\alpha {\dot A}'_\beta\}
\eqno{(4-25)}$$

Finally, take into account of all the above considerations, and
substitute eqs. (4-25) and (4-18) into the corresponding terms of
eq. (3-9) in which $e\rho\varphi$
being replaced by $e\rho (\varphi -\mu)$ as for a finite system,
we obtain an interesting form of the $Z$-generating functional for the
finite FQH system
$$Z=\int {\cal D}\theta_s^{bulk}{\cal D}\theta_s^{surf}\int {\cal D}\rho
\delta [{\cal F}[\rho, \rho_s^{bulk}+\rho_s^{surf};B]]$$
$$\cdot \exp~~i\left[\int d^3x \{-{\bf j}_s^{bulk}\cdot
{\bf A}'-{1\over m}\rho_s^{bulk}(e\varphi -\mu)\right. $$
$$\left.-{m \over 4\pi}\in_{\alpha \beta}A'_\alpha {\dot A'}_\beta
-V[\rho-{\bar \rho}\}+I_\Gamma [\theta_s]\right]
\eqno{(4-26)}$$
The surface action in eq. (4-26) $I_\Gamma$ has the form as
$$I_\Gamma [\theta_s]={1 \over 4\pi m}\int dt \oint_\Gamma dl
\{-n_\alpha \in_{\alpha \beta}
(\partial_\beta\theta_s{\dot \theta}_s
+\partial_\beta\theta_s^{bulk}{\dot \theta}_s
^{bulk})
$$
$$+{\tilde v}
_D(n_\alpha \in_{\alpha \beta}\partial_\beta\theta_s^{surf})^2
\}
\eqno{(4-27)}$$
where we have assumed the applied electric field ${\bf E}$
is parallel to the normal on the boundary and ${\tilde v}_D$ can be derived
from eq. (4-18) by applying eq. (4-8) as
$${\tilde v}_D=v_D/(1-2\pi\lambda^2 {\bar \rho}_s)
\eqno{(4-28)}$$
with $v_D=cE/B$.

In eq.(4-26), $\delta[{\cal F}[\rho,\rho_s;B]]$ is in fact a product
of $\delta-$functions
$$\delta[{\cal F}[\rho,\rho_s;B]]\equiv\prod_x
\delta[{\cal F}[\rho(x),\rho_s(x);B]],
\eqno{(4-29)}$$
where $\prod_x$ is the product over all the $2D$ spatial position
$x$'s and ${\cal F}[\rho(x),\rho_s(x);B]$ has exactly the
same expression as that of eq.(3-10) but picks
its value at the spatial points $x$.
Since the hardcore vortices can never coincide at the same spatial point,
we may regroup $\prod_x$ into two products as the following.
The first product, $\prod_{x \subset \Gamma}$, picks up those singular points
(attached with its nearest neighbouring regular points) at which  only the
surface vortices locate. Obviously, these "mini-islands"
(may or may not overlap) exist only in the
boundary layer region.
The second product, $\prod_{bulk}$, picks up all the other spatial points
in both the bulk interior and the remaining points in the boundary layer
region in which only the bulk vortices may locate. Therefore,
we may identify
$\rho_s(x)=\rho_s^{surf}(x)$ for those $\delta-$functions in the first product,
and
$\rho_s(x)=\rho_s^{bulk}(x)$ for those $\delta-$functions in the second
product.
We then have the following expression
$$\delta[{\cal F}[\rho,\rho_s;B]] \equiv\prod_{x\in\Gamma}
\delta[{\cal F}[\rho(x),\rho_s^{surf}(x);B]]\cdot
\delta[{\cal F}[\rho,\rho^{bulk};B]]
\eqno{(4-30)}$$
in which
$$\delta[{\cal F}[\rho,\rho^{bulk};B]]=\prod_{bulk}
\delta[{\cal F}[\rho(x),\rho^{bulk}(x);B]].
\eqno{(4-31)}$$
Keeping with the similar understanding, we may further separate the
integral measure of $\int{\cal D}\rho$ into two corresponding parts as
$$\int{\cal D}\rho=\int_{\Gamma}{\cal D}\rho\cdot
\int_{bulk}{\cal D}\rho
\eqno{(4-32)}$$
Now we introduce the notation
$$\tilde{{\cal D}}\theta^{surf}_s \equiv {\cal D}\theta^{surf}_s
\int_{\Gamma}{\cal D}\rho~\prod_{x\in\Gamma}
\delta[{\cal F}[\rho(x),\rho_s^{surf}(x);B]]
\eqno{(4-33)}$$
where $\int_{\Gamma}{\cal D}\rho~\prod_{x\in\Gamma}
\delta[{\cal F}[\rho(x),\rho_s^{surf}(x);B]]$ means to solve
$\rho(x)$ as the functional of $\rho^{surf}_s(x)$ from eq.(3-10)
in the boundary region. Taking into consideration of eqs.(4-30)$-\!-$(4-33),
the generating functional (4-26) can be put into the following form as
$$Z=\int_{\Gamma} \tilde{{\cal D}}\theta_s^{surf}
\int_{bulk} {\cal D}\theta_s^{bulk} {\cal D}\rho
\delta [{\cal F}[\rho, \rho_s^{bulk};B]]$$
$$\cdot \exp~i[\int d^3x
\{-{\bf j}_s^{bulk}\cdot {\bf A}'-{1\over m}\rho_s^{bulk}
(e\varphi -\mu) $$
$$-{m \over 4\pi}\in_{\alpha \beta}A'_\alpha {\dot A'}_\beta
-V[\rho-\bar{\rho}_s\}+I_\Gamma [\theta_s]]
\eqno{(4-34)}$$

For a finite system,
if the integration over ${\cal D}\rho \delta [{\cal F}[\rho,
\rho_s^{bulk};B]]$ has been taken into account, eq. (4-34) means that the
$Z$-generating functional for the FQH many electron system can be
equivalently described in terms of its vortex degrees of freedom
while the electrons can be understood as a background condensate.
The corresponding action can be divided into two parts: a bulk part and a
surface part. The bulk part has the
intuition that the vortices move in a dual gauge field $\bigtriangledown
\times {\bf A}'=-2\pi \rho^{bulk}$ and carry
the fractional statistics $(m)^{-1}$
with fractional charge $qe/m$. It can be interpreted as the action
for the next hierarchy. In particular,
when the system is exactly in a FQHE state of the first hierarchical level,
{\it i.e.}, $\rho_s^{bulk}=\theta_s^{bulk}=0$, we then have
$\theta_s=\theta_s^{surf}$, so that the surface action
$I_{\Gamma}[\theta_s] \to I_{\Gamma} [\theta_s^{surf}]$
will decouple from its bulk and describe an ensemble of independent
edge excitations with its propagation velocity ${\tilde v_D}=v_D$.
This is one of the interesting results drawn from our
approach with its description mainly based upon the constraint condition
eq. (3-10). We notice that if we solve $A'_\alpha$ in terms of $\rho^{bulk}$,
and
apply further eq. (3-10) for the $-{\bf j}_s^{bulk}\cdot {\bf A}'$ term,
we may find easily that the bulk action is formally rather similar to
that of [10, 11]. The action $
I_\Gamma [\theta_s^{surf}]$ in the FQH state has the form known as a
chiral boson action which is consistent also with those proposed in [14, 15].
What we have here is a unified description for a finite FQH system derived
from $ab~~initio$ analytically.

We stress further that if we perform a gauge transformation to the whole
action (4-26), it would
also produce a surface term which may cancel the surface term left previously
in section III. We will show the details in Appendix A.

As we have mentioned before, because $\theta_s(x)$ has only the isolated
singularities in the two dimensional plane, ${\cal D}\theta_s$ integrates
over only the space-time propagation of those singularities: the coordinates
of vortices. Therefore, it is not difficult to show that
$$\int {\cal D}\theta_s^{bulk}\exp~~i\int d^3x\{-{\bf j}_s^{bulk}\cdot {\bf A}'
-{m \over 4\pi}\in_{\alpha \beta}A'_\alpha \partial_0A'_\beta\}$$
$$=\sum_{N=1}^\infty \int \prod_{j=1}^N{\cal D}{\bf r'}_j(t)\exp~~i\{-\sum_j
{\dot {\bf r}'}_j\cdot {\bf A}'({\bf r'}_j(t))-{m \over 4\pi}\in_{\alpha
\beta}A'_\alpha \partial_0 A'_\beta\}
\eqno{(4-35)}$$
where ${\bf r'}_j(t)$ is the coordinate for the j-th bulk vortex.
We notice that, following from eq. (3-10), we always take the convention that
the vortices are counted as quasi-holes. This
identity makes the following fact become explicit. The bulk action for the
vortices in eq.(4-34) is essentially in a first quantization representation.
Moreover, it becomes clear that such an action again involves only terms
linear in the first order time
derivative of the vortex coordinates but no bilinear term. We may learn  from
the Dirac's algorithm immediately that once again we have
a system of vortices with ``zero kinetic energy'' which should be described
by the second class constraint. In fact, comparing eqs. (4-34) and (4-35)
with eq. (2-1), keeping again the understanding that the functional
integration over ${\cal D}\rho$ being carried through, we can realize that
the bulk action for the
vortices has a form almost the same as the original action for the electrons
in the LLL. Now it becomes also quite clear that the application of the
Dirac's quantization theory for the constrained systems to the overall
space-time propagation of the vortices in the form of eq. (4-35) provides
a field-theoretical background for treating these hierarchical vortices
( quasi-particles )
in FQHE which have only zero effective mass while the `` conventional ''
vortices often have
finite effective mass contributed by the massive constituting particles.

\bigskip
\bigskip

{\raggedright {\large \bf V Schematic Outline For The Higher Hierarchical
States And The
Corresponding Branches of Edge Excitations\\}}

Based on the above observations, we may apply the same procedure as those
for the electrons to introduce the second quantization representation for
the bulk vortices ( of the first hierarchy ). But there are certain
delicate differences which should be carefully treated as the following:
(i) Instead of the
vector potential ${\bf A}$ which couples to the electron velocity and has a
constant curl, $\bigtriangledown \times {\bf A}=B$ as the applied
magnetic field, we have now a vector potential ${\bf A}'$ for the bulk action
of the vortices which plays a similar role but has a curl,
$\bigtriangledown \times {\bf A}'=-2\pi \rho$, depending on the dynamical
variable {\it via} the constraint equation ${\cal F}[\rho,\rho^{bulk}_s;B]=0$;
(ii) In the application of the Dirac quantization to the vortices in the first
quantization representation, we need the condition $[{\Pi'_\alpha}^i,
{\Pi'_\beta}^j]=-2\pi \in_{\alpha \beta}\delta_{ij}\rho \not=0$ to be
satisfied, where ${\Pi'_\alpha}^i$ has the same form as $\Pi^i_\alpha$ with
the corresponding quantities substituted by those for the vortices.
Since $\rho$ could be zero ( or singular ) only at the isolated locations for
the vortices, in the spirit of long wave length approximation, we may
reasonably
take the approximation as $\rho >0$ ( finite ). In fact, these singular
behaviors at the
vortex locations will disappear after its second quantization procedure being
completed;
(iii) Corresponding to the bosonization procedure for
the electrons in which we introduced a C-S gauge field with the statistical
index being odd integers $m$, we now introduce a C-S gauge field $a'_\mu$ with
the statistical index being even integers $2p$. This is because that the world
lines for the vortex ``particles'' are originated from the singularities of
the phase field $\theta_s$ of the bosonized electrons, so that they have to
have a periodic boundary condition at the $-\infty$ and $+\infty$ of the time
axis [24].
By such a `` bosonization '' of the vortices, the newly introduced `` C-S ''
gauge field satisfies the gauge constraint as
$$\in_{\alpha \beta}\partial_\alpha {a'}_\beta=4\pi p\rho^{bulk}_s
\eqno{(5-1)}$$
Comparing eq. (5-1) with eq. (3-4), we have
$$a'_\alpha =-2p \partial_\alpha \theta^{bulk}_s
\eqno{(5-2)}$$
which in fact has the same physics as eq. (2-12). But eq. (2-12) is for the
electrons while eq. (5-2) is for the vortices with one hierarchical level in
succession. Substituting eq. (5-2) into eq. (4-23), we have
$${A'}_\alpha =-{1 \over {m\lambda^2 B}}{A_\alpha}^{em}+{1 \over 2pm}
{a'}_\alpha
\eqno{(5-3)}$$
This is a relation between the dual field and the new ``C-S'' field.

Taking into account of all the above considerations, introducing the
`` bosonized '' wave field $\Phi'$ for the bulk
vortices, and running over almost
exactly the same procedure as those for the electron case given  in the
section II, we may introduce the second quantization representation for the
vortex part of the the $Z$-generating functional (4-34). Consequently, it
can be transformed into the following form as
$$Z=\int\tilde{{\cal D}}\theta^{surf}_s
 {\cal D}\rho {\cal D}\Phi' {\cal D}{\Phi'}^+{\cal D}a'_\mu\delta
[[{\cal F}[\rho, \rho^{bulk}_s;B]]\delta
[{\tilde\Pi'}\Phi']\delta [{\Phi'}^+{\tilde {\Pi'}}^+]$$
$$\cdot \exp~i[\int d^3x \{{\Phi'}^+(i{\partial \over \partial t}-{1 \over m}
(e\varphi-\mu) -{a'}_0)\Phi'-V'[\rho_s]$$
$$+{1 \over 8p\pi} (
  2 a_0^{'} \in_{\alpha \beta} \partial_\alpha a^{'}_\beta
-   \in_{\alpha \beta} a_\alpha^{'} \partial_0 a^{'}_\beta 	)
-{1 \over {16\pi p^2 m}}\in_{\alpha \beta}{a'}_\alpha
\partial_0 {a'}_\beta\}+I_\Gamma [\theta_s ]]
\eqno{(5-4)}$$
where we have also substituted eq. (5-3) into eq. (4-34) and notice that the
first term on the r.h.s. of eq. (5-3) would not contribute to the C-S term
in eq. (4-34). Since $\theta_s=\theta^{bulk}_s+\theta^{surf}_s,$ we understand
that the dynamical variable $\theta_s$ for the surface action
has its bulk part
being now defined in the second quantization representation
while its surface part
being not. We notice further that in eq.(5-4) and the
following, except $\theta^{surf}_s$, all the second quantized dynamical
variables as well as their functional integration measure, such as $\rho$,
$\Phi'$ and ${\Phi'}^+$ etc.
are of bulk degrees of freedom,
and we would keep such understanding but ignore
the ``bulk'' sup- or subscripts for convenience.
Separating the modulus part of $\Phi'$ from its
phase part by writing $\Phi'={\sqrt {\rho_s}}e^{i\theta^{'}}$
with $\theta^{'}=\theta^{'}_r + \theta^{'}_s $, absorbing the regular part
of phase variable $\theta^{'}_r$ into $a_\mu^{'}$ (see Appendix A)
and then
integrating over ${a'}_0$, $a'_1$, $a'_2$ and $\rho$ in the $Z$-generating
functional (5-4) as what we did for the electrons in the section III, it
becomes
$$Z=\int \tilde{{\cal D}}\theta^{surf}_s
{\cal D}\rho_s{\cal D}{\theta'}_s\delta [{\cal F}'[\rho_s, \rho'_s;B]]
\exp~i[\int d^3x \{-\rho_s{\dot \theta'}_s-{1 \over m}\rho_s(e\varphi-\mu) $$
$$-{1 \over {16p^2 \pi}}({1 \over m}+2p)\in_{\alpha \beta}{a'}_\alpha
\partial_0 {a'}_\beta -V'[\rho_s]\}+I_\Gamma [\theta_s]]
\eqno{(5-5)}$$
with $V'[\rho_s]=V[(\rho_s-{\bar\rho}_s)]/m]$ and
$${\cal F}'[\rho_s, \rho'_s;B]\equiv {1 \over 2}\bigtriangledown^2\ln \rho_s -
{1 \over m\lambda^2}+2\pi \rho_s ({ 1\over m}+ 2p)+2\pi \rho'_s=0
\eqno{(5-6)}$$
where $a'_\alpha $ is the solution of eq. (5-1) associated
with an
appropriated gauge fixing condition which is determined again by the
constraint ${\hat\Pi'}\Phi'=0$ and its complex conjugate. ( see the
corresponding eqs.(3-7) and (3-8), especially (3-8) ). And
$\rho_s$, the density of the vortices, is the modulus of the vortex wave
field which
now is in the second quantization
representation,
while
$\theta'_s$ is the singular part for the conjugated phase
field which describes
the isolated `` vortices '' for the next ( higher ) hierarchical level
with its density having the expression as
$$\rho'_s={1 \over 2\pi}\in_{\alpha \beta}\partial_\alpha \partial_\beta
\theta'_s   \eqno{(5-7)}$$
These ``vortices'' has the intuition as `` new quasiholes'' on the ``old
quasihole'' condensate so that they are essentially electron-like excitations
in nature.
We may further solve ${\bar \rho}_s $ from eq. (5-6) with the consideration
of eq. (5-7) as
$${\bar \rho}_s={1 \over {2\pi (1+2pm) \lambda^2}}-{1 \over {m^{-1}+2p}}
{\bar \rho}'_s
\eqno{(5-8)}$$
In the above derivations, we have
carried out the path integral for ${\cal D}\rho$ so that the constraint
equation (3-10) ${\cal F}[\rho, \rho_s;B]=0$ is understood being always
satisfied and the ingredient of the constraint (3-10) has been now
transmitted into eq. (5-6). If we divide eq. (3-10) by $2\pi m$, eq. (5-6)
by $2\pi (m^{-1}+2p)$ and
then compare themselves each other, we may find that instead of $m^{-1}$ for
the vortices of the first hierarchical level, the charge unit of the vortices
of the
second hierarchical level becomes $-(1+2pm)^{-1}$. Correspondingly, the
statistics index
also changes from $m^{-1}$ to $-(m^{-1}+2p)^{-1}$.

We may further separate one more surface part of the action in eq. (5-5) from
the bulk in sense of the next hierarchical level. We may work
along exactly the same line as those of the electrons from eqs. (4-1)
to (4-28).
Since we have also the current conservation of the vortices: $\partial_0\rho'_s
+\partial_\alpha {j'}_s^\alpha =0$, and especially instead of eq. (3-10), we
have now the constraint equation eq. (5-6), therefore,
by noticing the correspondence as
$\rho\to -\rho_s$, $e\to -e/m$, $\theta_s\to -\theta'_s$ and the C-S
factor
$$m\to -({1\over m}+2p )
\eqno{(5-9)}$$
we may split $\rho'_s$ into $\rho_s ^{'surf}+ \rho_s^{'bulk}$,
$\theta'_s$ into $\theta_s^{'surf} + \theta_s^{'bulk}$ and
follow the same line as those of eqs. (4-1) to (4-10), and derive
$$({e \over m}\varphi -\mu')=-{e \over m}{\bf E}\cdot \delta {\bf r}'
\eqno{(5-10)}$$
with
$$\delta r_\alpha ={1 \over {2\pi (m^{-1}+2p ){\bar \rho}_s}}
[\in_{\alpha \beta}\partial_\beta {\theta'_s}^{surf} ]
\eqno{(5-11)}$$
in the boundary layer $x\subset \Gamma'$.
In repeating such a processing, we have an interesting question that whether
the ``boundary'' for the second hierarchical level $\Gamma'$ coincides
the boundary of the first hierarchical level $\Gamma.$
Formally, the FQH system
should have only one unique boundary on which all the surface integrals
for the system should be defined, {\it i.e.}, $\Gamma'=\Gamma$. But
intuitively,
as it has been already carefully discussed in the previous section,
the boundary
$\Gamma$ carries a sort of  ripple-like edge waves with an amplitude
of order of $\delta {\bf r}$.
It can be equivalently described in terms of the surface
vortices in sense of the first hierarchical level which are spreaded over a
surface region of depth $\sim\delta r$ and form a boundary layer.
We separated the surface degrees of freedom from those of the bulk in such
a way that the latter covers not only the whole region of the bulk interior
of the 2D FQH system but also the boundary layer in sense of those surface
vortices with its nearest regular neighbourhood being excluded.
This is the
basic physics of the boundary $\Gamma$, based upon which we introduced further
the surface measure $\tilde{{\cal D}}\theta^{surf}_s$ of eq.(4-33) and
the bulk measure in eq.(4-32). Following the same intuition, the boundary
$\Gamma'$ is in fact the boundary of the bulk region of
the first hierarchical level. It should be a rippling region with a depth
of $\delta r'$ but accommodates inside the
bulk region in a rather complicated way.
In other words, we could imagine that these two successive boundary layers
permeate into each other heavily, and we would like to say that
it is of the `` strong coupling limit ''.
We may imagine an opposite limiting case: all the surface vortices of the
second
hierarchical level, which is essentially the origin of the surface rippling
of the boundary $\Gamma^{'}$, distribute inside the boundary layer $\Gamma$
and form a layer as $\Gamma^{'}$.
We may have consequently the boundary layer
$\Gamma^{'}$ accommodates inside the
boundary layer $\Gamma$ with a sharp separation, $i.e.$,
up to the second hierarchical level, the FQH system has two
successive boundary regions with the outer boundary being $\Gamma$ while the
inner one being $\Gamma'$.
We say that  is of the `` weak coupling limit ''.
After the physics of the two coexisting boundaries being clarified as above,
corresponding to eqs. (4-11) and (4-12),
we have that, in the boundary layer $\Gamma^{'}$
$$\int_{x\subset \Gamma'}d^2x {\rho'_s}^{surf}=0
\eqno{(5-12)}$$
and
$${\bf j'}^{bulk}_s \cdot {\bf n'} |_{x\subset \Gamma'} = 0
\eqno{(5-13)}$$
where ${\bf n'}$ is the normal of the boundary $\Gamma ' $.
Keeping with such an understanding, we may process
further as follows.

Solving eq. (5-6) for $\rho_s$ and splitting then $\rho'_s$ into ${\rho'_s}^
{surf} +{\rho'_s}^{bulk}$,
we substitute it into the second term of the action in eq.(5-5).
We would like to keep the ${\rho'_s}^{bulk}$ term to be survived
and perform a partial integration for the remaining terms. This is in fact
the same procedure as done in eqs. (4-13)-(4-18) but with one hierarchical
level higher.
As the result, the term involving the applied electric field in action
eq. (5-5) can then be transformed into the following expression as
$$-{1 \over m}\int d^2x dt \rho_s (e\varphi -\mu)
={1 \over {1+2p m}}\int d^2x dt {\rho'_s}^{bulk}(e\varphi -\mu)$$
$$-{eE \over {2m (2\pi(m^{-1}+2p))^2 {\bar \rho}_s}}
\int dt\oint_{\Gamma'}
 dl (n_\alpha \in_{\alpha \beta}\partial_\beta {\theta'_s}^{surf})^2
\eqno{(5-14)}$$
On the meanwhile, we solve $a'_\alpha$ from eq. (5-1) and then utilize
eq. (5-6), we derive
$$a'_\alpha ={2mp \over {1+ 2mp}}\left[{1 \over m\lambda^2 B}A_\alpha ^{em}
-\partial_\alpha \theta'_s\right]
\eqno{(5-15)}$$
In the above equation we ignore the $\bigtriangledown^2\ln\rho_s$ term with the
same understanding as those for $\bigtriangledown^2\ln \rho$ in the previous
sections. And we introduce further a dual field ${\bf A}''$ for the new
bulk system
which is the correspondent of ${\bf A}'$ introduced by eq. (4-23)
$$A''_\alpha={1 \over {m^{-1}+2p}}[{1 \over m\lambda^2 B}A^{em}_\alpha
-\partial_\alpha {\theta'_s}^{bulk}]
\eqno{(5-16)}$$
with
$$\in_{\alpha \beta}\partial_\alpha {A''}_\beta =2\pi {\rho_s}^{bulk}
\eqno{(5-17)}$$
Then applying almost the same procedure as those from eq. (4-19) to
eq. (4-25) correspondingly,
the first as well as the third term of the action in eq. (5-5)
can be transformed into the following form as
$$\int d^2x dt [-\rho_s {\dot \theta'_s}-{1 \over {16 \pi p^2}}({1 \over m}
+2p)\in_{\alpha \beta}a'_\alpha {\dot a}'_\beta]$$
$$={1 \over 4\pi (m^{-1}+ 2p)}\int dt \oint_{\Gamma'} dl n_\alpha \in_{\alpha
\beta} (
 \partial_\beta \theta'_s {\dot \theta'}_s
+ \partial_\beta {\theta'_s}^{bulk}{{\dot \theta}_s}^{'bulk})$$
$$+\int d^2x dt
\left[-{\bf A}''\cdot {\bf j'}_s^{bulk} +{1 \over
4\pi} ({1 \over m}+2p)A''_\alpha \in_{\alpha \beta}{\dot A''}_\beta\right]
\eqno{(5-18)}$$
In deriving eq. (5-18), we notice that
the expression for the $\rho'_s$, eq.(5-7), has a formal sign
difference with that of eq.(3-4), therefore the corresponding expression
for ${\bf j}'_s$ should also has a sign difference with that of eq.(3-11)
formally.
Substituting eqs. (5-14) and (5-18) into eq. (5-5)
and applying further those arguments as well as treatments similar to that
of eqs.(4-29) to (4-34),
the $Z$-generating functional can be put into a new form as
$$Z=\int_{\Gamma}\tilde{{\cal D}}\theta^{surf}_s\int_{\Gamma'}\tilde{{\cal D}}
\theta'^{surf}_s \int_{bulk} {\cal D}
\rho_s {\cal D} \theta'_s\delta[{\cal F}'[\rho_s,\rho'^{bulk}_s;B]]$$
$$\cdot \exp~i\left[\int d^3x\{-{\bf j}'^{bulk}_s\cdot {\bf A}''+{1 \over
{1+2pm}}\rho'^{bulk}_s(e\varphi-\mu)-V'[\rho_s]\right.$$
$$\left.+{1 \over 4\pi}
({1 \over m}+2p)\in_{\alpha \beta}{A''}_\alpha \partial_0{A''}_\beta\} +
I_\Gamma [\theta_s]+I'_\Gamma [\theta'_s]\right]
\eqno{(5-19)}$$
with the additional surface action as
$$I'_\Gamma [\theta'_s]={1 \over 4\pi (m^{-1}+2p)}\int dt \oint_{\Gamma'} dl
\{ n_\alpha \in_{\alpha \beta}
  ( \partial_\beta\theta'_s {\dot \theta}'_s $$
$$
+\partial_\beta{\theta'_s}^{bulk}{{\dot \theta}_s}
^{'bulk}) -{\tilde
v'}_D(n_\alpha \in_{\alpha \beta}\partial_\beta {\theta_s}^{'surf})^2\}
\eqno{(5-20)}$$
where the drift velocity for the new edge excitations is now
$${\tilde v'}_D
=v_D/(1+2\pi m\lambda^2{\bar \rho}'_s).
\eqno{(5-21)}$$
For now we practiced our scheme once again that the bulk action for the
vortices eq. (4-34) can be also divided into two parts:
a surface part may describe one more branch of edge excitations, while the
remaining bulk part is exactly for the third hierarchical states. Both of them
have their forms almost the same as those given in eqs. (4-34) and (4-27),
and the only difference is that we have now the statistics parameter changed
from $-m$ to $m^{-1}+2p $ and the fractional charge changed from
$e/m$ to $-e/(1+2pm)$. Especially, noticing the sign difference between
eq. (3-4) and eq. (5-7), the surface action $I'_{\Gamma'}
[\theta'_s]$ of eq.(5-20) has consistently an additional global minus
sign compare
to $I_{\Gamma}[\theta_s]$ of eq.(4-27). The fact that these signs change
from one to the next reflects the hole-particle nature for the vortices
of different hierarchical levels which depends actually on our convention
that we keep the vortex particles as quasiholes for each hierarchical level.

For a homogeneous system with ${\bar \rho}'_s$ being equal to zero,
it means that the system is now lying exactly on the second hierarchical
FQHE filling, {\it i.e.}, we have a condensate for both electrons and vortices.
Then the constraints eq. (3-10),
${\cal F}[\rho,\rho_s;B]=0$, and eq. (5-6), ${\cal F}'[\rho_s, \rho'_s;
B]=0$, will give the expression for filling factor $\nu$ as
$$\nu ={1 \over \displaystyle m+
{\strut 1\over \displaystyle 2p}}
\eqno{(5-22)}$$
For the system having isolated vortices on the condensate of the second
hierarchical level, then
${\cal F}'[\rho_s, \rho'_s;B]=0$ will provide the corresponding vortex
equation with each vortex carrying a fractional charge as $(1+2pm)^{-1}$.
It becomes so obvious that our approach does provide a dynamical description
for these massless vortices for whole hierarchical scheme.

One more interesting
question is for the surface actions as we derived now two surface actions
co-existing in a FQH system at the second hierarchical level. For the second
one, eq. (5-20), we have the
understanding that $\theta'_s={\theta'_s}^{bulk}+{\theta'_s}^{surf}$ where
the ${\theta'_s}^{bulk}$ is contributed by the ${\rho'_s}^{bulk}$ while
${\theta'_s}^{surf}$ is contributed by the ${\rho'_s}^{surf}$. If the FQH
system is precisely on the second hierarchical level
with ${\rho'_s}^{bulk}={\theta'_s}^{bulk}=0$ so that
we have $\theta'_s={\theta'_s}^{surf}$ then the surface action eq. (5-20)
will be decoupled from the bulk as $I_{\Gamma'}[\theta'_s]\to I_{\Gamma'}
[{\theta'_s}^{surf}]$ and on the meanwhile, its drift velocity
${\tilde v}^{'}_D$
becomes $v_D$. But on the other hand, due to the boundary $\Gamma'$ accomodates
inside the boundary $\Gamma$, the ${\rho'_s}^{surf}$ and ${\theta'_s}^{surf}$
should contribute in principle to the ${\theta_s}^{bulk}$ variable defined
on $\Gamma$.Therefore, if we split
$\rho'_s$ into its bulk and surface part in eq. (5-6)
with the condition $\rho'^{bulk}_s=0$, we have
$$\rho_s={1 \over {2\pi \lambda^2(1+2pm)}}-{1\over
{m^{-1}+2p}}{\rho'_s}^{surf}
\eqno{(5-23)}$$
In eq.(5-23), $\rho_s$ actually satisfies eq.(4-5) since we had
ignore the ``bulk''-superscript for the second quantized $\rho_s$ after
(including) eq.(5-4). Moreover,$\rho'^{surf}_s$ satisfies an equation
of the same form as that of eq.(5-8) but with $\rho'_s$ and $\theta'_s$
substituted by $\rho'^{surf}_s$ and $\theta'^{surf}_s$ respectively.
Then we may solve $\partial_{\alpha}\theta^{surf}_s$ from eq.(5-23) as
$$\partial_\alpha \theta^{bulk}_s
={1 \over {m^{-1}+2p}}\left[\partial_\alpha \theta'^{surf}_s
-{1\over m\lambda^2 B}{A_\alpha }^{em}\right]
\eqno{(5-24)}$$
up to a trivial curl free 2-dimensional vector.
On the other hand, we may also express
$\theta^{bulk}_s$
directly in terms of $\rho_s$ which is entirely equivalent to eq.(4-5),
$$\theta^{bulk}_s=\int d^2x Im \ln ({\bar z}-{\bar z}')\rho_s (z')$$
subsequently, we have
$${\dot \theta}^{bulk}_s=\int d^2x Im \ln ({\bar z}-{\bar z}'){\dot \rho}_s(z')
\eqno{(5-25)}$$
For the $\theta'^{surf}_s,$ we should have similar equations followed
from eq.(5-7) with the condition $\rho'^{bulk}_s=0,$ these are
$$\theta'^{surf}_s=-\int~d^2 x~Im~ln(\bar{z}-\bar{z}')\rho'^{surf}_s,$$
and
$$\dot{\theta'}^{surf}_s=-\int~d^2 x~Im~ln(\bar{z}-\bar{z}')
\dot{\rho'}^{surf}_s.
\eqno{(5-26)}$$
By utilizing further eq. (5-23) again, we may show that
$${\dot \theta}^{bulk}_s={1 \over {m^{-1}+2p}}{\dot \theta}'^{surf}_s
\eqno{(5-27)}$$
The underlying physics could be understood as follows: due to the further
condensation of the vortices on the first hierarchical level, the singular
behavior
for the `` boundary '' vortices preserves and transmits itself into the
singular behavior for the vortices of the next hierarchical level {\it via}
the constraint equation (5-6) or eq.(5-23). Substituting
eqs. (5-24) and (5-27) into the surface action eq. (4-27), it becomes
$$I_\Gamma[\theta_s]\to I_\Gamma [\theta^{surf}_s,\theta'^{surf}_s]$$
$$={1 \over {4\pi m }}\int
dt \oint_\Gamma dl\{-[n_\alpha \in_{\alpha \beta}\partial_\beta
(\theta^{sur}_s+{1\over{m^{-1}+2p}}\theta'^{surf}_s)](\dot{\theta}^{surf}_s
+{1\over{m^{-1}+2p}}\dot{\theta}'^{surf}_s)$$
$$-{1 \over (m^{-1}+2p)^2}(n_\alpha \in_{\alpha \beta}\partial_\beta
{\theta'_s}^{surf}){{\dot \theta}_s}^{'surf}
+{\tilde v}_D(n_\alpha \in_{\alpha \beta}\partial_\beta
{\theta_s}^{surf})^2\}
\eqno{(5-28)}$$
\\
Eq.(5-28) contains only surface variables
$\theta^{surf}_s$ and $\theta'^{surf}_s$
so that they
decouple also from the bulk as long as the system is on the second
FQH hierarchy: $\rho'^{bulk}_s=0.$
But the two branches of edge excitations described by
$\theta^{surf}_s,\theta'^{surf}_s$
will formally couple to each other as shown by the explicit expressions
for actions $I_{\Gamma}[\theta^{surf}_s,\theta'^{surf}_s]$ and
$I'_{\Gamma'}[\theta'^{surf}_s]$ as eqs.(5-20) and (5-28) respectively.
In the weak coupling limit, {\it i.e.}, the two boundary layer $\Gamma$ and
$\Gamma'$ being sharply separated, due to ${\theta'_s}^{surf}$ has its source
${\rho'_s}^{surf}$ being nonzero only strictly inside the boundary layer
$\Gamma$, we may show in the Appendix B that ${\theta'_s}^{surf}$ will not
contribute to eq. (5-28). It would be then simplified to the following form as
$$I_\Gamma[\theta_s]\to$$
$$ I_\Gamma [\theta^{surf}_s]={1 \over {4\pi m }}\int
dt \oint_\Gamma dl\{-(n_\alpha \in_{\alpha \beta}\partial_\beta
\theta^{sur}_s)\dot{\theta}^{surf}_s$$
$$+{\tilde v}_D(n_\alpha \in_{\alpha \beta}\partial_\beta
{\theta_s}^{surf})^2\}
\eqno{(5-29)}$$
so that the two branches of edge excitations will further decouple
into two independent edge excitations. Associated with the action eq.(5-20)
in which ${\theta'_s}^{bulk}$ being now set to be zero,
$\theta'^{surf}_s$ describes one branch of edge excitation propagation
along boundary $\Gamma'$ with drift velocity $v_D$; while $\theta^{surf}_s$,
associated with the action (5-29), describes one another branch of
edge wave with the propagation velocity $\tilde{v}_D.$
The interesting point is that the latter
would has a different drift velocity
from that of $I_{\Gamma'}[\theta_s^{'surf}]$.
Following from eq. (5-8), we have now
${\bar \rho}_s =(2\pi \lambda^2 (1+ 2mp))^{-1}$ which is nomore zero
for the second FQHE hierarchical level. By
substituting it into eq.(4-28) we derive then
$${\tilde v}_D= v_D (1+{1 \over 2mp})
\eqno{(5-30)}$$
This is a rather interesting result that
we derived the analytical expressions for the propagation velocities of the
edge excitations which are different for its different branches. We expect
it could be checked by certain properly  designed experiment.

So far, we derived the corresponding edge excitations for the
second hierarchical level and the bulk action for the `` vortex '' of the
third hierarchical level
in which the `` vortex current '' would couple to a new `` C-S '' gauge field
as
$-{\bf j'}_s\cdot {\bf A}''$ with a C-S action $(4\pi)^{-1}(m^{-1}+2p)\in_{
\alpha \beta} A''_\alpha
{\dot A}''$. Now it is sufficiently convincing that by repeating the procedure
developed above, we arrive a complete description for the FQH system
that,
based upon a careful consideration of the LLL constraint,
the action incorporated with the constraint can be
transformed from one hierarchical state to the next in an almost
universal form, and the n-th hierarchical state can be viewed as
n branches of interacting edge excitations coupled to a (n-th) bulk
vortices system. In particular, only at the hierarchical filling of
the FQHE, these branches of edge state excitation will decouple from
bulk and bear the main physics of the FQHE state.

We would summarize further the
analytical
expressions for propagation velocities
of the edge excitations
hierarchically
as the following.
The statistics index $\kappa_n$ for the n-th hierarchical level has the
expression as
$$\kappa_n={1 \over {\kappa_{n-1}+2p_{n-1}}}
\eqno{(5-31)}$$
where $\kappa_{n-1}$ is the corresponding index for the (n-1)-th hierarchical
level with $\kappa_1=1/m$ and $p_{n-1}$ is an integer.
Then, the fractional charge for the vortices on the
(n-1)-th hierarchical states can be expressed as $e/m_n$ with
$$m_n=\prod^n_{l=1}\kappa_l^{-1}
\eqno{(5-32)}$$
in which we have $m_1=m$. And the vortex density for the (n-1)-th hierarchical
states can be expressed as
$$\rho^{(n-1)}={1 \over {2\pi \lambda^2m_n}}-\kappa_n\rho^{(n)}
\eqno{(5-33)}$$
with $\rho^{(n=0)}=\rho$. If the FQH system is on the N-th hierarchical
filling,
we have $\rho^{(N)}=0$, and the filling $\nu$ can be expressed as
$$\nu={1 \over m}[1-\kappa_1\kappa_2(1-\kappa_2\kappa_3(\cdots(1-\kappa_{N-1}
\kappa_N)\cdots ))]
\eqno{(5-34)}$$
If we substitute eq. (5-31) successively into eq. (5-34), it coincides
Haldane-Halperin expression [4] precisely. With the above notations, we
can show that the n branches of edge excitations for the n-th hierarchical
level have the general expressions as
$$v^{(j)}_D={v_D \over {1 -2\pi \lambda^2m_{j-1}\rho^{(j)}}}
\eqno{(5-35)}$$
with $j=1,\cdots, n$ and $v_D=cE/B$. In case of the FQH system being
on the N-th
hierarchical filling, {\i.e.}, $\rho^{(N)}=0$, we have then the hierarchical
expression for the drift velocities of the edge excitations as
$$v^{(1)}={v_D \over {1-\kappa_1\kappa_2(1-\kappa_2\kappa_3(\cdots (1-\kappa
_{N-1}\kappa_N)\cdots))}}={v_D \over m\nu}$$
$$v^{(2)}_D={v_D \over {1-\kappa_2\kappa_3(1-\kappa_3\kappa_4(\cdots (1-
\kappa_{N-1}\kappa_N)\cdots))}}$$
$$v^{(N-1)}_D={v_D \over {1-\kappa_{N-1}\kappa_N}}$$
$$v^{(N)}_D=v_D
\eqno{(5-36)}$$
We derive eq. (5-36) by substituting eqs. (5-33), (5-32) and (5-34) into
eq. (5-35).

\bigskip
\bigskip

{\raggedright{\large \bf VI SUMMARY AND DISCUSSIONS\\}}

In summary, our whole discussion is essentially based upon two basic
observations as follows.
The first is that since the vortices for any hierarchical level
( including the bosonized electrons ) have all their actions having only
terms linear in the vortex velocities, therefore, the Dirac algorithm
provides a highlight guiding line so that
we could have a unified treatment for the dynamics of the quasi-particles
in the FQH system. The second is that, in association with the constraint
for the LLL, a careful treatment of the partial
integrations in the actions for the finite FQH system may separate the
surface degrees of freedom from the bulk
which makes a proper description for the dynamics
of the edge excitations being possible.
What we have succeeded in this paper is
mainly that we derive not only
the expressions for the bulk actions as well
as the equations for the fractionally charged quasi-particles of each
hierarchical state,
but also the expressions of the actions, and subsequently the propagation
velocities, for the
associated branches of edge excitations analytically.
(We notify that, since the edge excitations are essentially a sort of
rippling wave of the boundary of an incompressible liquid, we,
as a primary study, ignored the effect of Coulomb interactions among the
surface vortices at the hierarchy filling.)  Especially,
we show that the branches of edge excitations can be decoupled from the
bulk only at the hierarchical fillings in the context of C-S field theory
approach. What we have found is that the
constraint equation, which can be transmitted from one hierarchical level to
the next, plays a central role in the whole formulation not only
for the bulk but also for the boundary.
We hope that the calculated
expressions for the propagation velocities of
the edge excitations could be checked experimentally.

\bigskip

{\raggedright{\large \bf ACKNOWLEDGEMENT\\}}
\bigskip

One of the authors (Z.B.S.) would like to thank Profs. L.N. Chang, D.H. Lee,
B. Sakita, S.C. Zhang for very useful discussions, especially he likes to
thank B. Sakita for his kind advisement and encouragement.
The authors would like also to thank
Drs. Y.X.Chen and S.Qin for useful discussions.
This work is partially supported by the NSFC, ITP-CAS and the CCAST.

\bigskip

{\raggedright{\large \bf APPENDIX A:\\}}

In the section III we have absorbed the regular functional $\theta_r$ in the
C-S gauge field $a_\mu$. This could be realized by performing a gauge
transformation $ a_\mu\to a_\mu-\partial_\mu \theta_r$
in eq. (3-6) and it gives
$$\int d^2x dt (-\rho{\dot \theta}_s-\rho{\dot \theta}_r -e\rho\varphi -\rho
a_0-V-{1 \over 2\pi m}a_0\in_{\alpha \beta}\partial_\alpha a_\beta
+{1 \over 4\pi m}\in_{\alpha \beta}a_\alpha {\dot a}_\beta)$$
$$=\int d^2x dt [-\rho {\dot \theta}_s -e\rho \varphi -\rho a_0 -V$$
$$-{1\over 2\pi m}(a_0 -{\dot \theta}_r)\in_{\alpha \beta}\partial_\alpha
(a_\beta-\partial_\beta \theta_r )+{1 \over 4\pi m}\in_{\alpha \beta}(a_\alpha
-\partial_\alpha \theta_r)({\dot a}_\beta -\partial_0\partial_\beta \theta_r)]
\eqno{(A-1)}$$
Utilizing the regular behavior of $\theta_r$: $\in_{\mu \nu \lambda}\partial_
\nu\partial_\lambda \theta_r=0$, and considering further that a term of total
time
derivative in the Lagrangian will give a zero contribution since the bosonized
system is periodic at $t=\pm \infty$, the r.h.s. of eq. (A1) can be transformed
into the following form by simple algebraic manipulations,
$$\int d^2x dt [-\rho {\dot \theta}_s-e\rho\varphi -\rho a_0-{1 \over 2\pi m}
a_0\in_{\alpha \beta}\partial_\alpha a_\beta +{1 \over 4\pi m}\in_
{\alpha \beta}a_\alpha {\dot a}_\beta -V]+{\cal K}_\Gamma[\theta_r,a]
\eqno{(A-2)}$$
with ${\cal K}_\Gamma[\theta_r,a]$ having the expression as
$${\cal K}_\Gamma[\theta_r,a]={1 \over 2\pi m}
\int dt \oint_\Gamma dl n_\alpha \in_{\alpha \beta} a_\beta {\dot\theta}_r-{1
\over 4\pi m}\int dt \oint_\Gamma dl n_\alpha \in_{\alpha \beta}\partial_\beta
\theta_r {\dot\theta}_r
\eqno{(A-3)}$$
In fact, ${\cal K}_\Gamma[\theta_r,a]$ is the right term which had been
forgotten tentatively in section III, especially in eq. (3-9).

On the other hand, the $\theta_s$ as well as ${\bf A}'$ dependent parts of the
action in eq. (4-26) have the following form
$${\cal L}[\theta_s, A'_\alpha]\equiv {1 \over 4\pi m}{\tilde v}_D\int dt
\oint_\Gamma dl (n_\alpha \in_{\alpha \beta} \partial_\beta \theta_s^{surf})^2
-{1 \over 4\pi m}\int dt \oint dl n_\alpha \in_{\alpha \beta}
( \partial_\beta
\theta_s^{bulk}{\dot \theta}_s^{bulk}$$
$$+ \partial_ \beta \theta_s{\dot \theta}_s )
-\int d^2x dt A'_\alpha j_{s,\alpha}^{bulk} -{m \over
4\pi}\int d^2x dt \in_{\alpha \beta}A'_\alpha {\dot A'}_\beta+{\cal K}_\Gamma
[\theta_r,a]
\eqno{(A-4)}$$
where we recovered the term ${\cal K}_\Gamma [a,\theta_r]$ and
introduced a notation ${\cal L}[\theta_s,A'_\alpha]$ for
convenience. If we perform further a gauge transformation as
$$\theta_s^{bulk} \to \theta_s^{bulk}-\theta_r$$
$$\theta_s^{surf} \to \theta_s^{surf}$$
$$A'_\alpha \to A'_\alpha +{1 \over m}\partial_\alpha \theta_r
\eqno{(A-5)}$$
for the action (A-4), {\it i.e.}, ${\cal L}[\theta_s,A'
_\alpha]\to {\cal L}[\theta_s-\theta_r,A'_\alpha+m^{-1}\partial_\alpha
\theta_r]$. The first term of eq. (A-4),
$(2\pi m)^{-1}{\tilde v}_D\int dt \oint_\Gamma dl~ (n_\alpha
\in_{\alpha \beta}\partial_\beta
\theta_s^{surf})^2$, is invariant under the gauge
transformation (A-5). Its second, third and fourth terms would be
transformed into
$$-{1 \over 4\pi m}\int dt \oint_\Gamma dl~n_\alpha
\in_{\alpha \beta}
[ \partial_\beta (\theta_s-\theta_r)({\dot \theta}_s-{\dot \theta}_r)
 + \partial_\beta (\theta_s^{bulk}-\theta_r)
       ({\dot \theta}_s^{bulk}-{\dot \theta}_r)]$$
$$-\int d^2x dt (A'_\alpha
+{1 \over m}\partial_\alpha \theta_r)j_{s,\alpha}^{bulk}
-{m\over 4\pi}\int d^2x dt
\in_{\alpha \beta}(A'_\alpha +{1 \over m}\partial_\alpha\theta_r)
({\dot A}'_\beta +{1\over m}\partial_0\partial_\beta \theta_r )
\eqno{(A-6)}$$
Substituting the gauge invariant expression for $j_{s,\alpha}^{bulk}$ as given
in eq. (3-11), and once again considering that $\theta_r$ satisfies
$\in_{\mu \nu \lambda}\partial_\nu \partial_\lambda\theta_r=0$ as well as the
fact that a total time derivative term in Lagrangian would give zero
contribution, we may transform eq. (A-6) into the following form {\it via}
step-by-step calculations:
$$-{1 \over 4\pi m}\int dt \oint_\Gamma dl
n_\alpha \in_{\alpha \beta}
( \partial_\beta \theta_s {\dot \theta}_s
 +\partial_\beta \theta_s^{bulk}{\dot \theta}_s^{bulk})$$
$$-\int d^2x dt A'_\alpha j_{s,\alpha}^{bulk} -{m \over 4\pi}\int d^2x
dt \in_{\alpha \beta}A'_\alpha {\dot A'}_\beta$$
$$+{1 \over 2\pi m}\int dt \oint_\Gamma dl~ n_\alpha \in_{\alpha \beta}
(\partial_\beta \theta_s){\dot \theta}_r-{1 \over 4\pi m}
\int dt \oint_\Gamma dl~ n_\alpha \in_{\alpha
\beta}{\dot \theta}_r \partial_\beta \theta_r
\eqno{(A-7)}$$
Moreover, substituting eq. (4-19) into eq. (A-3), the last term of eq. (A-4),
${\cal K }_\Gamma [\theta_r,a]$ would transform simultaneously into a form as
$${\cal K}_\Gamma[\theta_r,a]\to{\cal K}_\Gamma[\theta_r,a+\partial\theta_r]$$
$$=-{1 \over 2\pi m}\int dt\oint_\Gamma dl~n_\alpha \in_{\alpha \beta}
(\partial_\beta
\theta_s){\dot \theta}_r+{1 \over 4\pi m}\int dt \oint_\Gamma dl~n_\alpha \in_
{\alpha \beta}{\dot \theta}_r\partial_\beta\theta_r
\eqno{(A-8)}$$
Comparing eq. (A-8) with the last two terms of eq. (A-7), we
see that the surface terms in eq. (A-7) which is induced by the gauge
transformation eq. (A-5) are cancelled by ${\cal K}_\Gamma [\theta_r, a+
\partial \theta_r]$.

If we further take into account of the remaining term of the action in
eq. (4-26) $m^{-1}\rho_s^{bulk}(e\varphi
-\mu)$, with $\rho_s^{bulk}=-(2\pi)^{-1}\in_{\alpha \beta}\partial_\alpha
\partial_\beta \theta_s^{bulk}$, it is also invariant with respect to the gauge
transformation (A-5).
Consequently, the transformation eq. (A-5) indeed cancels the ${\cal
K}_\Gamma[a, \theta_r]$
term and keeps all the remaining terms have the form as in the text.

Alternatively, we may not cancel the surface term ${\cal K}_\Gamma
[a,\theta_r]$
at this stage and keep it to be remained as we process to the next hierarchical
level, $i.e.$, in eq.(5-4) we keep this additional term
${\cal K}_\Gamma [a,\theta_r]$ with $a_\alpha$ being defined as eq.(4-19)
{}.
Furthermore, similar to what we have done for the first hierarchical level,
there is one another
regular phase variable $\theta'_r$ in eq.(5-4) contributed by the vortex field
$\Phi'_s$ which arises from the second quantization representation of the
${\bf j'}^{bulk}_s \cdot {\bf A'}$ term in eq.(4-34) (or eq.(4-25)).

This $\theta'_r$ should
be absorbed into $a'_\mu$ {\it via} a transformation $a'_\mu
\to a'_\mu -\partial_\mu \theta'_r$
in the same way as those for eq.(3-6) ($i.e.$ eq.(A-1)).
Since $a'_\alpha =-2p \partial_\alpha
\theta_s^{bulk}$, $\theta_s^{bulk}$ should transform simultaneously as
$\theta_s^{bulk}\to \theta_s^{bulk}-(2p)^{-1}\theta'_r$ for consistency.
Therefore, all
those terms beside the ${\bf j}_s^{bulk}\cdot {\bf A}'$ in action (4-25)
$$-{1 \over 4\pi m}\int dt \oint_\Gamma dl~ n_\alpha
\in_{\alpha \beta}
( \partial_\beta \theta_s {\dot \theta}_s +
\partial_\beta \theta_s^{bulk}{\dot \theta}_s^{bulk})
-
{1 \over 16\pi p^2 m }
\int d^2x dt \in_{\alpha \beta} a'_\alpha {\dot a}'_\beta
\equiv {\cal R}
\eqno{(A-9)}$$
will transform accordingly as
$${\cal R}\to$$
$${\cal R}-{1 \over 4\pi mp}\int dt \oint_\Gamma dl~ n_\alpha \in_{\alpha
\beta}
a_\beta {\dot \theta}'_r-{1 \over 16\pi mp^2}\int dt \oint_\Gamma dl~n_\alpha
\in_{\alpha \beta}\partial_\beta \theta'_r {\dot \theta}'_r
\eqno{(A-10)}$$
The derivation from eq. (A-9) to eq. (A-10) actually is almost the same as
that from eq. (A-4) to eq. (A-7) with the ${\bf j}_s^{bulk}\cdot {\bf A}'$
term being kept away.
Correspondingly, noticing eq.(4-19),
the additional term  ${\cal K}_\Gamma [a,\theta_r]$ should transform
also into
$${\cal K}_\Gamma [a+{1 \over 2p}\partial \theta'_r, \theta_r]
={1 \over 4\pi m}\int dt \oint_\Gamma dl~ n_\alpha \in_{\alpha \beta}
\{
2{\dot \theta}_r a_\beta
+2{\dot \theta}_r ({1 \over 2p}\partial_\beta \theta'_r)
-{\dot \theta}_r \partial_\beta \theta_r
\}
\eqno{(A-11)}$$
If we set $\theta_r=(2p)^{-1}\theta'_r$, the ${\cal K}_\Gamma [a+(2p)^{-1}
\partial \theta'_s, \theta'_s]$ term will be cancelled exactly by the extra
terms
in eq. (A-10). On the meanwhile, the C-S term for the $a'_\mu$ field
with statistics index ${(8\pi p)}^{-1}$
will induce a new ${\cal K}'_\Gamma $ term ( due to absorbing the $\theta'_r$
variable )
leaving to the next higher hierarchical level. This part of discussion
indicates
that the additional surface term ${\cal K}_\Gamma [a,\theta_r]$ really does not
contributed to the dynamics of the next hierarchical level. Therefore, the
procedure in sections III and IV as well as the previous part of this appendix
that to cancel $\theta_r$ before going to the next hierarchical level is
reasonably correct.

\bigskip

{\raggedright {\large \bf APPENDIX B: }}

In section V, we derived the surface action of the boundary $\Gamma$ for the
system precisely on the FQH state of the second hierarchical level as
$$I_\Gamma [\theta_s^{surf},{\theta'_s}^{surf}]={1 \over 4\pi m}\int dt
\oint _\Gamma dl\{n_\alpha \in_{\alpha \beta}\partial_\beta \theta_s^{surf}
{\dot \theta}_s^{surf}-{\tilde v}_D(n_\alpha \in_{\alpha \beta}\partial_\beta
\theta_s^{surf})^2\}$$
$$-{1 \over {2\pi m(m^{-1}+2p)^2}}\int dt \oint_\Gamma dl~n_\alpha
\in_{\alpha\beta}\partial_\beta {\theta'_s}^{surf}{{\dot \theta}_s}^{'surf}$$
$$-{1 \over {2\pi m(m^{-1}+2p)}}  \int dt \oint_\Gamma dl~n_\alpha
\in_{\alpha \beta}\partial_\beta \theta_s^{surf}{{\dot \theta}_s}^{'surf}
\eqno{(B-1)}$$
It is straightforward to verify that eq. (B-1) is exactly identical to
eq. (5-28). In eq. (B-1), it is known from the sections IV and V that
$$-{1 \over 2\pi}\in_{\alpha \beta}\partial_\alpha \partial_\beta \theta_s
^{surf}=\rho_s^{surf}
\eqno{(B-2)}$$
$${1 \over 2\pi }\in_{\alpha \beta}\partial_\alpha \partial_\beta {\theta'_s}
^{surf}={\rho'_s}^{surf}
\eqno{(B-3)}$$
where $\rho_s^{surf}$ is nonzero only in the boundary layer $\Gamma$ while
${\rho'_s}^{surf}$ is nonzero only in the layer $\Gamma'$.

In the weak coupling limit, the boundary layer $\Gamma'$ is enclosed inside
the boundary layer $\Gamma$ with a sharp separation. It is
equivalently to say that the bundle of world lines for the surface
vortex particle
( described by $\rho'_s$ ) in $\Gamma'$ will never penetrate into the bundle
of the world lines of surface
vortex particles in $\Gamma$ ( although they are vortex
particles in sense of different hierarchical level ). Based upon such an
assumption ( approximation ), we will show in this appendix that the third and
fourth terms on the r.h.s. of eq. (B-1) have zero contribution.

Introduce
$${\rho'_s}^{surf} (x)=\sum_{j\subset \Gamma'}q'_j\delta^2 ({\bf x}-{\bf x}'_j
(t))
\eqno{(B-4)}$$
and
$${\rho_s}^{surf} (x)=\sum_{i\subset \Gamma}q_i\delta^2 ({\bf x}-{\bf x}_i
(t))
\eqno{(B-5)}$$
where $q_i$ and $q'_j$ are the vortex charge for the vortex particle $i$ and
$j$
respectively. Then we may solve ${\theta'_s}^{surf}$ and $\theta_s^{surf}$
from
eqs. (B-2) and (B-3) as
$${\theta'_s}^{surf}(x)=-\sum_{j\subset \Gamma'}q'_j Im\ln ({\bar z}-{\bar
z}'_j
(t))
\eqno{(B-6)}$$
$${\theta_s}^{surf}(x)=\sum_{i\subset \Gamma}q_i Im\ln ({\bar z}-{\bar z}_i
(t))
\eqno{(B-7)}$$
and subsequently,
$${{\dot \theta}_s}^{'surf}(x)=\sum_{j\subset \Gamma'}q'_j{{\dot x}_j}^
{'\alpha }(t)\partial_\alpha  Im\ln ({\bar z}-{\bar z}'_j
(t))
\eqno{(B-8)}$$
By applying eqs. (B-6), (B-7) and (B-8), the third and fourth terms in eq.
(B-1)
can be rewritten as
$${1 \over {2\pi m(m^{-1}+2p)^2}}\sum_{j\subset \Gamma'}\sum_{j'\subset
\Gamma'}q'_jq'_{j'}\int dt {{\dot x}_{j'}}^{'\gamma} (t) \cdot$$
$$\oint _\Gamma dl~n_\alpha
\in_{\alpha \beta}\partial_\beta Im \ln ({\bar z}-{{\bar z}'_j}(t))\partial_
\gamma Im \ln ({\bar z}-{\bar z}'_{j'}(t))$$
$$
+{1 \over {2\pi m(m^{-1}+2p)}}\sum_{i\subset \Gamma}\sum_{j'\subset
\Gamma'}q_iq'_{j'}\int dt {{\dot x} _{j'} } ^{'\gamma} (t) \cdot$$
$$\oint _\Gamma dl~n_\alpha
\in_{\alpha \beta}\partial_\beta Im \ln ({\bar z}-{{\bar z}_i}(t))\partial_
\gamma Im \ln ({\bar z}-{\bar z}'_{j'}(t))
\eqno{(B-9)}$$
Utilizing
$$\oint_\Gamma dl~n_\alpha \in_{\alpha \beta}\partial_\beta=\oint_\Gamma
dl_\alpha \partial_\alpha =\oint _\Gamma (dz \partial_z +d{\bar z}\partial_
{\bar z})$$
and
$${\dot x}_\gamma \partial_\gamma ={\dot z}\partial_z+{\dot {\bar z}}\partial_
{\bar z},$$
eq. (B-9) becomes
$${1 \over {2\pi m(m^{-1}+2p)^2}}\sum_{j\subset \Gamma'}\sum_{j'\subset
\Gamma'}q'_jq'_{j'}\int dt \oint _\Gamma dl $$
$$\{dz \partial_z
Im \ln ({\bar z}-{{\bar z}'_j}(t)){\dot z}'_{j'}(t)\partial_z
Im \ln ({\bar z}-{\bar z}'_{j'}(t))$$
$$+d{\bar z} \partial_{\bar z}
Im \ln ({\bar z}-{{\bar z}'_j}(t))
{\dot {\bar z}}'_{j'}(t)\partial_{\bar z}
Im \ln ({\bar z}-{\bar z}'_{j'}(t))$$
$$+dz \partial_z
Im \ln ({\bar z}-{{\bar z}'_j}(t))
{\dot {\bar z}}'_{j'}(t)\partial_{\bar z}
Im \ln ({\bar z}-{\bar z}'_{j'}(t))$$
$$+d{\bar z} \partial_{\bar z}
Im \ln ({\bar z}-{{\bar z}'_j}(t)){\dot z}'_{j'}(t)\partial_z
Im \ln ({\bar z}-{\bar z}'_{j'}(t))\}$$
$$+{1 \over {2\pi m(m^{-1}+2p)^2}}\sum_{i\subset \Gamma}\sum_{j'\subset
\Gamma'}q_iq'_{j'}\int dt \oint _\Gamma dl $$
$$\{dz \partial_z
Im \ln ({\bar z}-{{\bar z}_i}(t)){\dot z}'_{j'}(t)\partial_z
Im \ln ({\bar z}-{\bar z}'_{j'}(t))$$
$$+d{\bar z} \partial_{\bar z}
Im \ln ({\bar z}-{{\bar z}_i}(t))
{\dot {\bar z}}'_{j'}(t)\partial_{\bar z}
Im \ln ({\bar z}-{\bar z}'_{j'}(t))$$
$$+dz \partial_z
Im \ln ({\bar z}-{{\bar z}_i}(t))
{\dot {\bar z}}'_{j'}(t)\partial_{\bar z}
Im \ln ({\bar z}-{\bar z}'_{j'}(t))$$
$$+d{\bar z} \partial_{\bar z}
Im \ln ({\bar z}-{{\bar z}_i}(t)){\dot z}'_{j'}(t)\partial_z
Im \ln ({\bar z}-{\bar z}'_{j'}(t))\}
\eqno{(B-10)}$$

We would like to discuss the eight group terms of eq. (B-10) term by term. If
we take the derivatives to the imaginary part of the $ln $ function, any of
the first group term of eq. (B-10) would be proportional to
$$\oint _\Gamma dz{1 \over {(z-z'_j(t))(z-z'_{j'}(t))}}=0
\eqno{(B-11)}$$
where we have utilized the fact that, as what we have assumed, ${\bf x}'_j(t)$,
${\bf x}'_{j'}(t)$ always stay inside the $\Gamma$. With the similar arguments,
we can show easily that the second, fifth and sixth group terms are also equal
to zero. If we take a partial integration with respect to $dz \partial_z$ for
any of the third group term of eq. (B-10), it would transform into a form
proportional to
$$\oint_\Gamma
Im \ln ({\bar z}-{\bar z}'_j(t))  {\dot {\bar z}}'_{j'}(t)
\partial_z \partial_{\bar z}
Im \ln ({\bar z}-{\bar z}'_{j'}(t))$$
$$=\oint_\Gamma dz Im \ln ({\bar z}-{\bar z}'_j(t)){\dot {\bar z}}_{j'}(t)
(-i\pi )\delta ^2 ({\bf x}-{\bf x}'_{j'} (t))
\eqno{(B-12)}$$
where we have made use of the identities
$$(\partial_z \partial_{\bar z}-\partial_{\bar z}\partial_z)Im\ln ({\bar z}-
{\bar z}_{j'}(t))= - 2\pi i\delta^2 ({\bf x}-{\bf x}_{j'}(t)),$$
$$(\partial_z \partial_{\bar z}+\partial_{\bar z}\partial_z)Im\ln ({\bar z}-
{\bar z}_{j'}(t))=0.$$
Since ${\bf x}_{j'}(t)$'s stay always inside the $\Gamma $ while ${\bf x}$ is
in the $\Gamma$, the $\delta^2 ({\bf x}-{\bf x}'_j(t))$ in eq. (B-12)
should always
take the value zero. As a result the third group term of eq. (B-10) has only
zero contribution. By applying the similar arguments, we may show also that the
fourth, seventh and eighth group terms of eq. (B-10) do not contribute too.

Consequently, in the weak coupling limit we have shown in this appendix
that eq. (B-1),
{\it i.e.}, eq. (5-28) can be simplified into a form as eq. (5-29)
$$I_\Gamma [\theta_s^{surf}]={1 \over 4\pi m}\int dt \oint_\Gamma dl\{-
n_\alpha \in_{\alpha \beta}\partial_\beta \theta_s^{surf}{\dot \theta}_s^{surf}
+{\tilde v}_D(n_\alpha \in_{\alpha \beta}\partial_\beta \theta_s^{surf})^2\}
\eqno{(B-13)}$$
which indeed decoupled form the ${\theta'_s}^{surf}$ right on the filling
of the second hierarchical level.

\bigskip

{\raggedright {\large \bf REFERENCES \\}}
\bigskip
\begin{enumerate}

\item {D.C. Tsui, H.L. Stormer, A.C. Gossard, {\it Phys. Rev. Lett.} {\bf 48},
1559(1982)}
\item {R. Prange, S.M. Girvin, {\it The Quantum Hall Effect} Springer
Verlag, (1990)}
\item {R.B. Laughlin, {\it Phys. Rev. Lett.} {\bf 50}, 1395(1983)}
\item {F.D.M. Haldane, {\it Phys. Rev. Lett.} {\bf 51}, 605(1983); B.I.
Halperin, {\it Phys. Rev. Lett.} {\bf 52}, 1583(1984)}
\item {S.M. Girvin, A.H. MacDonald, P.M. Platzman, {\it Phys. Rev. Lett.}
 {\bf 54}, 581(1985); {\it Phys. Rev.} {\bf B33}, 2481(1986)}
\item {S. Kivelson, C. Kallin, D.P. Arovas, J.R. Schrieffer, {\it Phys. Rev.
Lett.} {\bf 56}, 873(1986); G. Baskaran, {\it Phys. Rev. Lett.} {\bf 56},
2716(1986)}
\item {S.M. Girvin, A.H. MacDonald, {\it Phys. Rev. Lett.} {\bf 56},
1252(1987)}
\item {Also see, E.H. Rezayi, F.D.M. Haldane, {\it Phys. Rev. Lett.} {\bf 61},
1985(1988); X.C. Xie, Song He, S. Das Sarma, {\it Phys. Rev. Lett.} {\bf 66},
389(1991)}
\item {F. Wilczek, {\it Phys. Rev. Lett.} {\bf 49}, 957(1982); F. Wilczek,
A. Zee, {\it Phys. Rev. Lett.} {\bf 51}, 2250(1983); D.P. Arovas, J.R.
Schrieffer, F. Wilczek, {\it Phys. Rev. Lett.} {\bf 53}, 722 (1984)}
\item {S.C. Zhang, H. Hansson, S. Kivelson, {\it Phys. Rev. Lett.} {\bf 62},
82(1989); {\bf 62}, 980(1989); M.P.H. Fisher, D.H. Lee, {\it Phys. Rev. Lett.}
{\bf 63}, 903(1989); D.H. Lee, S.C. Zhang, {\it Phys. Rev. Lett.} {\bf 66},
1220(1991)}
\item { D.H. Lee, {\it Int. J. Mod. Phys.} {\bf B5}, 1695(1991);
S.C. Zhang, {\it Int. J. Mod. Phys.} {\bf B6}, 25(1992) }
\item {N. Read, {\it Phys. Rev. Lett.} {\bf 62}, 86(1989)}
\item {X.G. Wen, {\it Phys. Rev.} {\bf B41}, 12838(1990); {\it Phys. Rev.
Lett.}
{\bf 64}, 2206(1990); {\it Int. J. Mod. Phys.} {\bf B6}, 1711(1992)}
\item { B. Blok, X.G. Wen, {\it Phys. Rev.} {\bf B42}, 8133(1990);
D.H. Lee, X.G. Wen, {\it Phys. Rev. Lett.} {\bf 66}, 1765(1991)}
\item { M. Stone, {\it Ann. Phys.} (N.Y.) {\bf 207}, 38(1991)}
\item {C.W.J. Beenakker, {\it Phys. Rev. Lett.} {\bf 64}, 216 (1990);
A.H. MacDonald, {\it Phys. Rev. Lett.} {\bf 64}, 220 (1990)}
\item {X.G. Wen, {\it Phys. Rev.} {\bf B40}, 7387(1989); X.G. Wen, Q. Niu, {\it
Phys. Rev.} {\bf B41}, 9377(1990); X.G. Wen, A. Zee, {\it Phys. Rev.} {\bf
B44},
274(1991); {\it Phys. Rev. Lett.} {\bf 69}, 953(1992)}
\item {J.K. Jain, {\it Phys. Rev. Lett.} {\bf 63}, 199(1989); {\it Phys. Rev.}
{\bf B41}, 7653(1991); A. Lopez, E. Fradkin, {\it Phys. Rev.} {\bf B44}, 5246
(1991)}
\item {A short version of this paper,
preprint AS-ITP-92-44,
has been submitted for publication.}
\item {P.A.M. Dirac, {\it Lectures on Quantum Mechanics}, Belfer Graduate
School of Science, Yeshiva University, New York, (1964)}
\item {B. Sakita, D.N. Sheng, Z.B. Su, {\it Phys. Rev.} {\bf B44}, 11510(1991)}
\item {Zhong-Shui Ma, Zhao-Bin Su, {\it The Guiding Center Coordinate and
Constraint of
The Lowest Landau Level For The Planer Electron System}, to be published }
\item {R. Jackiw and So-Young Pi derived a similar equation but without the
applied magnetic field, {\it Phys. Rev. Lett.} {\bf 64}, 2969(1990)}
\item {Z.F. Ezawa, A. Iwazaki, {\it Phys. Rev.} {\bf B43}, 2637 (1991);
Zhong-Shui Ma, Zhao-Bin Su, {\it A 2+1 Dimensional Dual Chern-Simons Field
Approach For The Fractional Quantum Hall System}, to be published }

\vspace{30ex}

\end{enumerate}
\end{document}